\documentclass[sigconf]{acmart}
\AtBeginDocument{%
  }

\usepackage{placeins}
\usepackage{xcolor}
\usepackage{enumitem}
\usepackage{booktabs}
\usepackage{graphicx}
\usepackage{tabularx} 
\usepackage[prologue, table,xcdraw]{xcolor}
\usepackage{subcaption}
\usepackage{soul} 
\usepackage{ulem}

\begin{document}

\title{How Tech Workers Contend with Hazards of Humanlikeness in Generative AI}

\author{Mark D\'iaz}
\email{markdiaz@google.com}
\affiliation{%
  \institution{Google Research}
  \country{USA}
}

\author{Renee Shelby}
\email{reneeshelby@google.com}
\affiliation{%
  \institution{Google Research}
  \country{USA}
}

\author{Eric Corbett}
\affiliation{%
  \institution{Google Research}
  \country{USA}
}

\author{Andrew Smart}
\email{andrewsmart@google.com}
\affiliation{%
 \institution{Google Research}
 \country{USA}
}
\renewcommand{\shortauthors}{D\'iaz et al.}

\begin{abstract}
    Generative AI's humanlike qualities are driving its rapid adoption in professional domains. However, this anthropomorphic appeal raises concerns from HCI and responsible AI scholars about potential hazards and harms, such as overtrust in system outputs. To investigate how technology workers navigate these humanlike qualities and anticipate emergent harms, we conducted focus groups with 30 professionals across six job functions (ML engineering, product policy, UX research and design, product management, technology writing, and communications). Our findings reveal an unsettled knowledge environment surrounding humanlike generative AI, where workers' varying perspectives illuminate a range of potential risks for individuals, knowledge work fields, and society. We argue that workers require comprehensive support, including clearer conceptions of ``humanlikeness'' to effectively mitigate these risks. To aid in mitigation strategies, we provide a conceptual map articulating the identified hazards and their connection to conflated notions of ``humanlikeness.''
\end{abstract}

\maketitle

\section{Introduction}
\label{Introduction}
The rapid adoption of generative AI (genAI) by technology companies and other businesses has introduced new interfaces for acquiring knowledge, problem-solving, and sociality~\cite{skjuve2024people}, presenting novel challenges for developers and researchers. GenAI technologies, particularly for text or verbal dialogue, often exhibit characteristics perceived as \textit{humanlike}~\cite{abercrombie2023mirages}. These characteristics include illusory cognitive abilities (e.g., reasoning and problem-solving), sophisticated communication skills (e.g., generating fluid, conversational language)~\cite{shanahan2024talking}, and even the simulation of emotional expression~\cite{character2024, chai2024, miAI2024, replika2024}. While  interpreted as ``humanlike,'' these capabilities function through computational algorithms obscured by user interfaces~\cite{burrell2016, suchman2023uncontroversial} and can be enhanced through the humanlikeness of the interface itself or its interaction modality.

The perceived ``humanlikeness'' of genAI, driven by technological advancements and contextual experience, raises important responsibility and safety questions regarding its design, interpretation, and integration into sociotechnical systems. Anthropomorphism can be a means for people to interpret and make sense of a technology~\cite{muller2004, lupton1997} and is shaped by sociocultural factors~\cite{waytz2010social}. Without careful risk mitigation, however, this phenomenon can lead to interaction-based sociotechnical harms~\cite{shelby2023, weidinger2022, weidinger2021ethical} as ``humanlike'' features can influence interactions and may ``prime users to interact with information'' in particular ways~\cite[p. 1070]{maeda2024}. Although all technologies carry risks that must be managed, accepted industry standards for developing humanlike AI features for identifying and mitigating the hazards they introduce do not yet exist. These challenges necessitate understanding how tech workers using, developing, or who are otherwise impacted by these technologies understand the implications of humanlike qualities that ``afford''~\cite{davis2023} certain human-AI relations and perceptions.

While attributing human qualities to non-human entities (i.e., anthropomorphism~\cite{epley2007seeing, epley2008we}) is a well-established psychosocial phenomenon in HCI~\cite{reeves1996media, muller2004, nass2000machines, moon1996}, genAI's contemporary  sophistication brings renewed urgency to understanding the interplay of genAI features and user perceptions. Unlike traditional interfaces and chatbots, genAI mimics human behaviors with remarkable fidelity, influencing user perceptions~\cite{abercrombie-etal-2021-alexa}, level of trust~\cite{inie2024, jensen2021, CULLEY2013577}, emotional attachment~\cite{berney2024care}, disclosure~\cite{kim2022}, expectations~\cite{maeda2024}, and perceived quality of interactions~\cite{kaate2024, PELAU2021106855, ARAUJO2018183}. This raises important questions about tech workers' perceptions of humanlikeness hazards, particularly for those at the vanguard of using, implementing, and otherwise responding to these computational systems with limited guidance: 
\label{RQs}
\begin{enumerate}[leftmargin=1.4cm,itemsep=0.15cm,label=\bfseries RQ\arabic*:]
    \item  What are technology workers' perceptions of the development and use of humanlike genAI? 
    \item What potential hazards for their roles, fields, and society are most salient to technology workers?
\end{enumerate}
We address our research questions by examining the perspectives of 30 tech workers across six job functions-- ML engineering, product policy, UX research and design, product management, technology writing, and communications-- in small to enterprise-level companies. In our inquiry, we distinguish between humanlikeness and anthropomorphism, which have been conflated in prior work,  and choose to frame our questions around humanlikeness for two reasons: (1) humanlike design does not guarantee anthropomorphism, but  still carries  risks, such as inducing unrealistic user expectations of systems \cite{fink2012anthropomorphism}; (2) we are interested in worker perceptions beyond anthropomorphism, including risk assessments and their general impressions of how \textit{others} perceive and use humanlike genAI.

We focus on tech workers as they occupy a critical position in the genAI landscape~\cite{boyarskaya2020, cooper2022, green2020}, being the ground-zero users responsible for integrating genAI into workflows and products, as well as navigating its use by others. While facing challenges similar  to other workers newly interacting with genAI, they are among the first to engage with genAI due to the tech industry's early adoption and influence downstream use and impacts through their product design, policy, and development decisions. 
The oft-hidden design and governance decisions made by tech workers carry far-reaching implications for users and publics~\cite{star1999layers, suchman1995making}, constituting a central facet of the ``human work of AI''~\cite[p. 3]{fox2023}: how technologies are shaped by practices of work. 

Our findings illuminate epistemic and practical challenges workers face navigating humanlike AI in organizational processes and specifying the relationship between anthropomorphism and its concomitant impacts on various hazards and harms. We found tech workers operate in an unsettled knowledge environment~\cite{nelkin1992controversy}--drawing from \textit{positional and partial knowledge} to fill  gaps left by missing guidance and policy that is needed to identify hazards that humanlike genAI poses to people, knowledge work fields, and society. Their knowledge reveals multiple, sometimes competing conceptions of humanlikeness (Section \ref{humanlikeness}), while showing they articulate a broad range of hazards associated with different humanlike features despite navigating with limited information (Section \ref{lowinformation}). Synthesizing these insights, we provide a  conceptual map detailing the relationship between humanlikeness and hazards in genAI (Section \ref{discussion}). We conclude by advocating for greater specificity in defining and operationalizing humanlikeness goals in AI development and offer insights into how the fields of HCI and responsible AI (RAI) can intervene to better guide genAI development and use to mitigate potential harms.
\section{Related Work}
Our work builds on rich literatures within CSCW and adjacent communities on humanlike design features, sociotechnical harms, and the practices of workers who build algorithmic systems.

\subsection{Humanlikeness in GenAI}
\label{humanlikeness-in-genai}

In discussing humanlike genAI, we distinguish between  ``humanlikeness,'' referring to design choices that emulate human characteristics or behaviors, and ``anthropomorphism,''  referring to the psychological phenomenon of how people perceive technological agents, regardless of  explicit design intent. 

\subsubsection{Humanlike Design Features}
We define ``humanlikeness'' as  incorporating  human attributes into AI systems to cultivate natural and engaging user interactions. Such features can reflect a wide range of familiar human communication patterns~\cite{seeger2021texting}, including: \textit{human identity cues}, such as sociodemographic markers \cite{cowell2005manipulation, riedl2011trusting} and visual representations \cite{berry2005evaluating, gong2008social}; \textit{linguistic cues}, encompassing social chit-chat \cite{chattaraman2019should, bickmore2005social}, emotional expressions \cite{de2016almost}, pronoun usage \cite{chattaraman2019should, sah2015effects}, and contextual responsiveness \cite{knijnenburg2016inferring,  schuetzler2014facilitating}; and \textit{behavioral cues}, such as emoticons \cite{derks2008emoticons}, temporal cues \cite{feine2019taxonomy, gnewuch2018faster}, turn-taking gestures \cite{de2016almost}, and embodied movements \cite{schneiders2021effect}. Developers implement humanlike design features to improve user interaction and foster ``natural'' interactions, drawing on the notion that users prefer systems mirroring human communication~\cite{chandra2022or}. Examples range  widely, from animated agents featuring 3D, humanlike appearances and synchronized speech~\cite{liew2022anthropomorphizing} to  subtle  gestural cues, such as a login form shaking to indicate an incorrect password \cite{valenzuela2017implications}, or a blinking light that evokes the appearance of an eye \cite{baraka2018mobile}. In virtual agents, including chatbots and voice assistants, humanlike design cues influence how users perceive the agent and its social presence \cite{ARAUJO2018183}.

\subsubsection{Effects of Humanlike Design Features}
While  humanlike design aims to create intuitive interfaces, its incorporation yields mixed impacts. Linguistic cues in chatbots can improve social presence, emotional connection, and purchase intent \cite{tsai2021chatbots, toader2019effect, konya2023someone}, while visual cues, particularly those resembling humans or animals, may foster emotional bonds and social interaction \cite{qiu2009evaluating, christoforakos2023technology, maeda2024}. However, effectiveness hinges on contextual factors (e.g., \cite{pawlik2021design, miesler2012product, catrambone2019anthropomorphic, inie2024}) and correct calibration \cite{chandra2022or}. Cultural preferences for humanlikeness vary \cite{epley2007seeing}, and misusing these cues can lead to disappointment, frustration, or eeriness \cite{rajaobelina2021creepiness, glikson2020human}, especially when the humanlike form suggests capabilities the technology lacks, creating unrealistic expectations \cite{fink2012anthropomorphism}. Excessive humanlikeness can distract users and reduce their sense of control \cite{laestadius2024too}. Given how humanlikeness eases interactions, this raises concerns for AI-enabled manipulation and coercion \cite{akbulut2024all}, particularly in conversational AI, where trust and attachment can be exploited \cite{inie2024}.  Creating  emotional connections with AI also raises concerns about privacy, autonomy, and the potential dehumanization of service provision \cite{maeda2024}. Ultimately, appropriate calibration requires careful consideration of potential benefits and harms within a specific context. However, precisely measuring and characterizing the harms of humanlike design features remains a challenge.

\subsubsection{Anthropomorphism: User Perception}  Anthropomorphism—— attributing  humanlikeness to non-human entities~\cite{epley2007seeing}——is a long-standing area of computing research focused on human tendencies to perceive and interact with technology as they would with other humans~\cite{nass1993anthropomorphism, gong2008social, lee2010more, don1992anthropomorphism}. Anthropomorphism is one way users `construct’ or make sense of technology~\cite{suchman1993categories, Hales1994} and is an inherently social process. HCI researchers have interrogated factors influencing anthropomorphism, including underlying technological capabilities~\cite{reeves1996media} and observable humanlike cues~\cite{ha2021}, as well as its impact on user perceptions and behaviors~\cite{li2021,bi2023}. Early work demonstrated parallels between human-human and human-computer interactions, including phenomena like reciprocal sharing~\cite{moon2000intimate}, ingroup and outgroup social dynamics~\cite{nass1996can}, and social stereotyping~\cite{nass1997machines}; however, people apply unique social scripts to human-AI interaction~\cite{gambino2020building}. More recently, scholars have explored its influence on genAI interactions, including user trust~\cite{waytz2014mind}, behavior~\cite{adam2021ai}, and system adoption~\cite{sheehan2020customer}. These studies emphasize the importance of interrogating how anthropomorphism shapes human-AI interaction, particularly  the implications of parasocial human-AI relationships~\cite{maeda2024, brandtzaeg2022, pentina2023}. While features and effects of anthropomorphism are delineated, how tech professionals conceptualize humanlikeness and its risks— amidst the rapid evolution of genAI— remains a critical area for investigation.

\subsection{Harms and Hazards of Humanlike GenAI Features}

All technologies carry inherent hazards with the potential for harm, necessitating identification and mitigation as part of an organization's safety culture~\cite{leveson2016engineering, cooper2000}. These harms are \textit{sociotechnical}, arising from the contextual interplay of technical system components and societal power dynamics~\cite{shelby2023}. Researchers have begun identifying potential harms of humanlike genAI, including: breakdowns in human connection~\cite{woodruff2024}, maladaptive human-AI interaction~\cite{holstein2020conceptual}, overtrust relative to  actual reliability~\cite{seymour2021, maeda2024, mehrotra2024, manzini2024should}, deskilling~\cite{woodruff2024}, and macroeconomic impacts on knowledge work~\cite{eloundou2023, briggs2023}, among others. The OECD~\cite[p. 3]{oecd2022} further emphasizes the risk of ``mass persuasion and manipulation ... as a result of AI anthropomorphism.'' Consequently,  mapping anthropomorphism-related harms is highlighted as a necessary component of broader genAI safety~\cite{cheng2024one}. 

\subsubsection{Knowledge Gaps Hinder Responsible Development.} While humanlike features encompass communicative, visual, and aural aspects of interaction, as well as content~\cite{pfeuffer2019},  no standardized evaluation approaches exist~\cite{li2022anthropomorphism}. Current work ties certain system features to some interaction-based outcomes, such as trust~\cite{moussawi2021effect}, which has intuitive ties to sociotechnical harms related to manipulation and fraud, or \textit{emotional dependence}~\cite{laestadius2024too} which can lead to mental health harms~\cite{skjuve2021my, vaidyam2019chatbots}; yet, this understanding is limited.
The field currently lacks a comprehensive understanding of the relationships between humanlike features in genAI, anthropomorphism, and potential harms, challenging the determination of effective interventions~\cite{cheng2024one}. Critically, not all humanlike design features exert the same influence on anthropomorphism~\cite{gabriel2024ethics}. Additionally, anthropomorphism can occur even without explicit humanlike design—for instance, perceiving a ``sad'' or ``happy'' face in car headlights. For these reasons, our investigation focuses on hazards of humanlike genAI, regardless of specific design intent. Psychological research suggests anthropomorphism is an automatic process~\cite{guthrie1995faces, mitchell1997anthropomorphism}, occurring even when systems lack obvious humanlike cues~\cite{WANG2017334} or when individuals are explicitly informed that an interface is distinct from the machine. This indicates  digital literacy alone does not preclude the phenomenon~\cite{nass1997machines}.  

However, anthropomorphism is not inherently pathological or harmful. It can offer benefits, such as facilitating socially appropriate interaction~\cite{zlotowski2015anthropomorphism} and trust calibration~\cite{de2016almost}, holds sociohistorical value~\cite{jensen2013}, and appears across various cultural forms~\cite{guthrie1995faces, geerdts2016real}. Thus, it can exert both positive and negative influences on human-AI interaction.

\subsubsection{``Humanlike''  Features as Sociotechnical Hazards.} In contrast to calls to minimize anthropomorphism in AI \cite{abercrombie2023mirages}, we frame anthropomorphism  as a \textit{hazard}: a condition that creates the potential for downstream harm. Similarly, we view humanlike features generally as potential contributors to both anthropomorphism and other hazards. Effective risk management necessitates identifying and controlling these hazards in both direct (e.g., a user interacting with a chatbot) and indirect human-AI interactions (e.g., someone reading genAI content published online). This can be achieved through control strategies implemented at societal, organizational, and worker levels~\cite{rasmussen1994risk}, encompassing law and policy, research-based best practices and standards, organizational policies and practices, and practices implemented by individual tech workers across roles. However, knowledge gaps persist throughout this control hierarchy, leaving individual workers to manage risks despite limited resources and supports. While existing HCI guidelines on human-AI interaction~\cite{amershi2019, chancellor2023toward} and humanlike design~\cite{abercrombie2023mirages} offer a starting point, they lack specific guidance on identifying and mitigating downstream hazards introduced by these features.

\subsection{Tech Workers Navigating Poor Information Environments in the Era of GenAI}
The rapid growth of the genAI market \cite{bloomberg2024} presents challenges for tech workers using and building humanlike genAI. Hype-driven tech cycles  (for critiques, see:~\cite{bartholomew2023, chomsky2023}) create sensationalized narratives about genAI~\cite{woodruff2024} and foster poor information environments for workers~\cite{vinsel2023}. This situation can  lead workers to rely heavily on tacit knowledge or outdated guidance, potentially resulting in design flaws that only emerge after deployment~\cite{scheuerman2024products}. Within hype-driven \textit{sociotechnical landscapes}~\cite{rip1998technological}, workers face increased difficulties making informed decisions about using and communicating about humanlike genAI~\cite{woodruff2024}.

Anticipating the social implications of a technology is  challenging, particularly as technological features often  afford interactions across a positive to negative continuum~\cite{schiff2020principles, maris2022}. 
Whether and to what degree a feature risks harm is contextual, necessitating careful analysis within the specific sociotechnical context (i.e., the product and its situated  use)~\cite{rismani2023}. Perspectives and beliefs towards the particular use of a genAI tool ground people's beliefs about potential harms and harm reduction opportunities~\cite{shelby2024}. However, specifying every detail of a sociotechnical system upfront is rarely possible, as vulnerabilities can arise over time, emerge from interaction between system components, or stem from dynamics within and beyond the immediate context of use~\cite{dobbe2024toward}. 

Knowledge gaps concerning the effects of humanlike genAI are exacerbated by the distributed nature of tech work, where workers focus on different aspects of system development and use~\cite{slota2023many}, often relying on \textit{situated knowledge}~\cite{suchman1987plans} specific to their role and context. Understanding how they engage in \textit{sensemaking}~\cite{weick1995sensemaking} within  poor ``information environments''~\cite{vinsel2023} is necessary, as their interpretations shape downstream design and policy decisions. 
Consequently, the choices workers across various job functions make carry significant weight. For example, humanizing language in public communications can exaggerate system capabilities \cite{placani2024anthropomorphism},  and design decisions, such as allowing chatbot customization, influences user behavior~\cite{bi2023}. Absent specific guidance and robust knowledge, workers may use and respond to humanlike genAI without fully understanding the potential implications of their decisions, thereby shaping user experiences and societal impacts in  unforeseen ways. Investigating tech worker perspectives and hazard perception can illuminate necessary interventions.

Ultimately, this work advances scholarship on humanlike genAI by 1) investigating perceptions of humanlikeness and its associated hazards through workers on the ground, 2) providing a conceptual map as a basis for further empirical study and foundational knowledge on humanlikeness hazards, and 3) extending prior work on RAI guidance to identify current gaps and pathways for supporting workers contending with the latest genAI developments.
\section{Methodology}
\label{method}

To understand technology workers' attitudes toward and experience with humanlike genAI (RQ1) and the potential hazards it introduces (RQ2), we conducted a series of focus groups. Focus groups allowed us to examine the shared knowledge of tech workers,  conceptualized here as a specific category of knowledge worker \cite{morgan1996focus, krueger2014focus}, providing insights into the field's current understanding of humanlikeness and perceived hazards.

\begin{table*}[t]
\centering
\begin{tabular} {l l r} 

 \toprule
\textbf{Job Role}                    				  & \textbf{ID}                     & \textbf{Total} \\ 
\midrule

Product Manager &  P2, P4, P14, P15, P16, P17, P18	      & 7      \\ 
Machine Learning Engineer  &       P5, P19, P20, P21, P22, P23		 &  6     \\ 
Tech Writer/Content Strategist  &  P1, P7, P27, P28, P29	   &  5     \\ 
Policy Analyst        &  P24, P25, P26   &   3    \\ 
UX Designer/Researcher                 &   P3, P9, P10, P11, P12, P13	  &   6    \\ 
Communications/Public Relations Manager &     P6, P8, P30   & 3  \\ 

\bottomrule
\end{tabular}
\caption{Breakdown of participants by job role. Participants represented a range of job functions common to the tech field.}
\label{main}

\end{table*}

\begin{table*}[t] 
    \centering
    \begin{subtable}[t]{0.45\textwidth} 
        \centering
        \caption{Organization Size}
        \label{tab:orgsize}
        \begin{tabular}{lr}
            \toprule
            \textbf{Organization Size} & \textbf{Count} \\
            \midrule
            < Self-employed/contractor & 4 \\
            < 200 employees & 9 \\
            < 1,000 employees & 3 \\
            < 5,000 employees & 3 \\
            5,000+ employees & 11 \\
            \bottomrule
        \end{tabular}
    \end{subtable}%
    \hfill 
    \begin{subtable}[t]{0.45\textwidth} 
        \centering
        \caption{Years in Field}
        \label{tab:years}
         \begin{tabular}{lr}
            \toprule
            \textbf{Years in Field} & \textbf{Count} \\
            \midrule
            <2 years & 3 \\ 
            2-5 years & 11 \\ 
            5-10 years & 10 \\ 
            10+ years & 6 \\ 
            \bottomrule
        \end{tabular}
    \end{subtable}
    \label{demos}
    \caption{Worker experience. Workers were employed across a range of organization sizes, including those who were self-employed or contractors up to large, multi-national companies. The majority of workers had between 2 and 10 years of experience in their field.} 
    \label{tab:demographics} 

\end{table*}

\subsection{Participant Recruitment} 
We recruited 30 U.S.-based adult tech workers across six job functions, with 5-6 participants per function: UX designers/researchers, technical writers and content strategists, engineers and research scientists, product managers, communications and public relations, and policy analysts (see Tables~\ref{main} and ~\ref{tab:demographics}).\footnote{While large tech companies typically feature Trust \& Safety teams focused on product policy and assessing development practices, smaller organizations may rely on  policy analysts.} We recruited general tech workers rather than only those building genAI products, because workers who use genAI in their role and those interacting with colleagues and end-users who, in turn, use genAI offer valuable perspectives on the broader risk landscape. Although only four worked on  model pre-/post-training and another five worked in organizations with first-party genAI product offerings, all possessed domain expertise relevant to emerging genAI impacts. Recruiting across roles enabled us to capture concerns beyond engineering and those rooted in cross-functional coordination.

We sourced participants from a recruiting partner (anonymized for review) experienced in recruiting tech workers. The roles served as windows into the views of tech workers and fostered collective sensemaking with professional peers; they were not intended to be exhaustive or for systematic comparison. We worked with the recruiting partner to identify  job functions typical for tech companies of various sizes and asked potential participants to indicate a job description  best aligning with their current role and to describe their job in their own words, rather than strictly screening out imperfect matches.
Participants had to be at least 18 years of age, live in the U.S. and self-report familiarity with genAI systems (e.g.,  ChatGPT, Gemini). The  protocol and  materials were approved by our legal and ethics review process. Participants received a gift card in thanks for their participation. 

\subsection{Focus Groups} 
We held one remote focus group per job role plus a mixed-role group, totaling  seven sessions, conducted in October-November 2023. The mixed-role focus group was held first, but as the discussion strayed from workers' direct experiences to adjacent topics (e.g., non-U.S. regulation), we proceeded with role-specific groups. We found that role-specific groups facilitated more focused discussions about common work tasks and shared challenges. We reached theoretical saturation before closing our initial round of data collection. Planning one group per role balanced breadth of inquiry with the depth necessary for rich qualitative analyses \cite{guest2017many} and  ensured a ‘thick description’ of the discourse, avoiding superficial analysis \cite{carlsen2011n}.

Focus groups began with an introduction to humanlikeness in non-humans generally, then in genAI, to ground participants in a common  understanding. We provided examples of technologies sometimes labeled ‘humanlike’ but offered no specific definition, inviting participants to share prior experiences and externalize their conceptions. We used the term ``humanlike’’ and briefly introduced ``anthropomorphic,’’  avoiding academic jargon that risked intimidating workers. The focus group instructions and introduction to humanlikeness can be found in Appendix \ref{appendix}. While we broadly introduced humanlikeness, discussion of genAI most frequently anchored on AI chat bots, given their current prevalence and public attention.

The groups were then presented with a series of provocations asking their agreement with statements related to humanlike genAI in their work context. The provocations were inspired by recent work discussing genAI advancements (Q1) \cite{bubeck2023sparks} that will change work due, in part, to humanlike capabilities (Q2, Q6) \cite{leaver2023chatgpt}, as well as  barriers to following responsible AI guidance \cite{madaio2024learning, heger2022understanding}, motivating inquiry into workers' state of knowledge and views on genAI hazards (Q3, Q4, Q5). 
The provocations were formulated  reductively to spur richer discussion where workers could clarify nuances. This format was used in a discussion activity facilitated during a CRAFT session held at FAccT 2023 and attended by the lead author \cite{wilkinson2023theories}. For each provocation, participants  voted on an initial stance (agree, disagree, agree with both, or agree with neither), before  deeper discussion. Participants were invited  to challenge the  options or  premises and were asked  to avoid sharing NDA-sensitive information or using specific product, person, or team names. Discussions centered on worker perceptions of humanlike genAI and its associated hazards, drawing from both their general genAI knowledge and work domain expertise. This included concerns related to genAI use among colleagues, by other companies, and product end users. 
\begin{enumerate}
    \item Humanlikeness in generative AI is fundamentally different from ``humanlikeness'' in other technologies (e.g., Siri, Clippy).
    \item The prevalence of ``humanlike'' generative AI (e.g., ChatGPT, Bard) has made my job easier.
    \item In my work context, ``humanlike'' generative AI introduces more or new risks compared with other technologies.
    \item I am confident about the considerations needed to make decisions regarding how to think about and/or develop ``humanlike'' genAI.
    \item In my work, there are clear guidelines for how I should design, assess, or think about ``humanlikeness'' in genAI.
    \item ``Humanlike'' generative AI will change the future of my work.
\end{enumerate}
Three to four researchers moderated each workshop to facilitate  discussion and take notes. All sessions lasted approximately two hours and were recorded.

\subsection{Data Analysis}
We transcribed recordings and analyzed data using a hybrid codebook thematic analysis (TA)  combining inductive and deductive approaches. Our process  followed five-stage codebook TA \cite{roberts2019attempting}.

\paragraph{Sourcing initial codes} 
\label{initialcodes}
Before data collection, we analyzed literature on 1) anthropomorphism and digital technology; 2) Science and Technology Studies (STS) and interpretive flexibility, and 3) genAI's impact  on knowledge workers. Existing literature on anthropomorphism and humanlike features informed our initial deductive codes; codes describing hazards, harms, and worker reactions to humanlike genAI were developed inductively, as these are emergent areas in the literature.

\paragraph{Developing initial codes}
All four researchers open-coded and memoed one transcript, then discussed to draft initial open codes. Then all authors coded the same second transcript, followed by  discussion leading to  a final high-level codebook covering workers’ humanlikeness definitions, reactions to genAI, and salient hazards. The  team then applied existing codes deductively and generated new codes inductively. We included supplementary humanlikeness definition codes from Waytz~\cite{waytz2010sees} (i.e., ``has a mind of its own,'' ``has free will,'' ``has intentions,'' ``can experience emotion,'' ``has consciousness'') to organize quotes implying \textit{anthropomorphizing}, as  workers did not explicitly discuss anthropomorphizing processes. 

\paragraph{Codebook design}
The research team collaboratively refined the codebook labels, definitions, and exclusions. It was segmented into three sections: how workers reason about and conceptualize genAI,  how they see humanlike genAI in relation to their work, and perceived challenges to workers, industry, and society. 

\paragraph{Codebook application}
The research team applied the codebook to transcripts using ATLAS.ti, creating additional memos on emergent themes. Each transcript was coded by two researchers. The authors held weekly meetings to reflexively discuss findings. We assembled hazard codes and discussed related humanlikeness characteristics, referencing  quote contents to  determine  explicit and implicit mentions.

\paragraph{Interpretation}
We exported quotes with code labels to an online whiteboard (Mural) for grouping and organizing. Here, we conducted additional thematic groupings of sub-themes by code groups and job role. Based on code mappings, we surfaced a conceptual map relating humanlikeness characteristics and downstream risks.

\subsection{Limitations}
This work carries the standard limitations of purposive samples~\cite{palinkas2015purposeful} and focus group dynamics (e.g., tendency for socially acceptable opinions or participant dominance). These tendencies, however, reveal normative values and practical ideologies~\cite{smithson2000using}. While the focus of this research is on tech workers broadly, the sample may not fully capture the range of perspectives in tech work. While this study offers rich insights into the perspectives of our participants, findings may not generalize broadly to all job functions.
Finally, the approach  prioritized gathering a breadth of risks at the expense of capturing depth in specific job role distinctions. Due to this design choice and confidentiality restrictions (NDAs), we did not capture  granular decisions and their subsequent product risk impacts.
\section{Findings}
\label{findings}
Humanlike genAI presents epistemic and practical challenges for workers' understanding of AI systems, partly stemming  from pressures to adopt genAI despite limited guidance. 
We find tech workers draw on both professional and lay experiences to articulate humanlikeness and sensemake about harms and hazards of humanlike genAI, even though discussions were grounded in professional contexts. On a backdrop of limited guidance, workers drew from their positional knowledge as users, developers, and non-user stakeholders in their job roles, revealing particular patterns in the hazards salient to each  role. We first detail workers' understanding of humanlikeness  (\ref{humanlikeness}),  before detailing the six hazards of humanlikeness (\ref{hazards}) and summarize perspective differences across  roles (\ref{positional-differences}).

\subsection{Multiple Conceptions of ``Humanlikeness’’ in Human-AI Interaction }
\label{humanlikeness}
A fundamental prerequisite for addressing humanlikeness hazards is developing targeted mitigation strategies. However, workers' conceptualizations of ``humanlike'' varied significantly, relating to interaction dynamics, perceived task sophistication, and individual anthropomorphizing tendencies. These varied conceptions highlight that ``humanlikeness'' is a multifaceted construct shaped by situated experiences and interpretation,  revealing conceptual barriers to aligning mitigation strategies.

\subsubsection{Interaction Dynamics} 
Interaction dynamics were central to tech workers' conceptions of ``humanlikeness.'' This included open-ended, multi-turn interactions, socially-appropriate affective responses to user inputs, and the influence of visual and physical embodiments on user behavior. They viewed such interactions as potentially ``anthropomorphic'',  affording new human-AI relations that ``brin[g] the user more connected towards the product itself,'' citing virtual reality avatars as an example (P18, Product Manager). 

Workers emphasized how genAI’s responsiveness and multi-turn interactions fundamentally shifted the nature of human-AI interaction. One UX worker (P13) noted, ``we are moving into a realm where we are [intuitively] interacting, not necessarily [explicitly] asking a certain question,'' while a product manager (P14) suggested  genAI's responsiveness makes it ``actually act like a human'' compared to earlier ML systems that simply ``mimic’’ humans. 

Key ``humanlike'' interaction dynamics included responsiveness to commands, the ability to tailor communication to users' tonal preferences, and adherence to social norms and niceties. One content manager (P7) described how iterative use of genAI chatbotsfosters a sense of collaboration and allows for dynamic, nuanced modifications impossible with traditional computer interfaces: 
\begin{quote}
...you can do something similar that you can do with other people ... like let's go back and forth on this and keep modifying something in small increments. You can ask it to change things in little ways that you didn't used to be able to do with a computer.
\end{quote}
Social niceties, like a chatbot apologizing for an incorrect response, were  noted aspects of ``humanlikeness.''\footnote{Some developers have made purposeful decisions not to have their chatbots apologize (e.g., Anthropic's \cite{anthropic2024} Claude)} 
Centrally, however, genAI's humanlikeness manifested as contextual awareness, particularly adapting to different communication styles and affective states. As one product manager (P29) described: ``I honestly think that... we get more comfortable with them [as users] as they become more human and more tangible and relatable.''

\subsubsection{Conversational Input} GenAI's ability to process conversational language was closely tied to perceived humanlikeness. Workers highlighted the significance of natural language input, contrasting it with technologies requiring unnatural or unintuitive query formulation (e.g., as with voice assistants and traditional web search). Responding to diverse, naturally phrased questions was  key, making genAI chatbots feel like ``a person'' answering  questions (P30, Public Relations).
Others echoed this shift: a communications worker (P7) emphasized the ``intuitiveness'' of ``language-based interaction'' while using genAI, while P12 in UX noted the move towards ``more natural conversation ... mov[ing] away from keywords and more towards a phrase or a concept.'' Conversational input thus enhanced perceived humanlikeness and fundamentally changed user engagement.

\subsubsection{Humanlike Outputs} GenAI outputs mirroring human language or utterances were significant markers of ``humanlikeness,'' including language fluidity, grammar quality, tone, and even errors  mimicking common human mistakes: ``It's so human on a small sentence to sentence level'' (P7, Content Manager). Outside conversational contexts (e.g., AI-generated documents),  workers noted genAI content feels humanlike in tautological ways: because it's written in a humanlike way: ``It's an inherent byproduct that is going to feel humanlike because those [training] mediums are written in a humanlike way'' (P13, UX). Interestingly,  realistic errors contributed, suggesting imperfections  paradoxically enhance perceived humanness. P7 noted certain comma errors made chatbots seem more human, suggesting learning from erroneous text:

\begin{quote}
Sometimes when ChatGPT makes errors... it almost makes it more human.  I've wondered if the text it learned on had errors. There's certain comma errors that it makes pretty consistently, but they're also really common comma errors that people make.
\end{quote}
The capacity for generating creative, ``original,'' and expressive outputs also contributed to perceptions of humanlikeness, contrasting  with older AI systems (e.g., rule-based ML) that lacked such capabilities. An engineer (P23) praised genAI's creative text generation for flyers he creates in volunteer work outside of his job as ``much more creative than I am.'' In this way, genAI outputs were deemed humanlike both for exceeding a human baseline, and for closely mimicking it, flaws included.

\subsubsection{Task Sophistication} 
The ability of genAI to perform tasks typically associated with humans was another key factor influencing perceptions of  humanlikeness, even if  system design was not overtly humanlike, such as Tesla's self-driving capabilities (P14, Product Manager). Less ``hard-coded'' and seemingly unexpected behaviors also contributed to a sense of humanlikeness and discoverability. An ML engineer (P23), referencing AlphaGo (a non-genAI system), described defying expectations as distinctly humanlike. Similarly, generating humorous content impressed some as a marker of complexity and humanlikeness:

\begin{quote}
I happen to think ChatGPT is hilarious. I have so many examples of just really funny things that it's come up with and that's kind of just a human aspect of it. It's ability to just synthesize massive amounts of data and somehow know what will come off as funny. (P21, Communications)
\end{quote}
Conversely, a UX designer questioned genAI's social awareness and competence, such as understanding paralinguistic communication,
\begin{quote}
AI is not aware of the five senses. It’s not looking at our face to see, `oooh I shouldn’t talk about this like that' or this answer is too verbose because someone looks like they are straining to read it. And so it doesn’t have all the [feedback mechanisms]. (P3, UX)
\end{quote}

\subsubsection{Anthropomorphizing} 
Lastly, discussions revealed workers' tendency to anthropomorphize even basic features, reflecting how people imbue systems with social meaning beyond actual capabilities or design intent \cite{nass1997machines}. Evidence included perceiving personalities in car headlights (P15, Product Manager), and the practice of naming work projects to foster connection. P15 explained how naming projects helps ``humanize the relationship between you and that product or service'' and gives them ``personality'' and you a ``better relationship.'' This reflects both explicit and implicit \textit{anthropomorphizing}. It is notable that anthropomorphism theory does not suggest that using anthropomorphizing language does not guarantee deep, existential beliefs, yet it has been shown to influence the social behavior of users and audiences \cite{morris2007metaphors}.

At the same time, complex, contextual factors not  tied to specific design features also contributed to perceptions of humanlikeness. P6, who works in communications for a medical technology company, described attributing emotions and morality to genAI based on its overall behavior, suggesting   inferences arise from multiple factors and social context: ``the way [genAI] mimics emotion, and seems to know-- not always gets it right-- but seems to have an idea of right from wrong and to really experience it.''
Similarly, a system's intended role influenced humanlike perceptions. A product manager (P4) described his mother's reliance on her voice assistant, where dependence fostered by humanlike qualities became apparent when connectivity failed: ``She is like `I can't have my machine to answer the questions that I need [when the wifi goes down].' And so...the anthropomorphic piece is there.''
This suggests social context and perceived purpose shape how users perceptions and use, regardless of actual capabilities.

\subsection{The Hazards of ``Humanlikeness''}
\label{hazards}
Beyond general development concerns, tech workers identified specific hazards related to a wide range of humanlike characteristics of genAI. These hazards were not mutually exclusive and often intersected, though the salience of individual hazards varied by job role (discussed further in \ref{standpoints}).

\subsubsection{The Hazard of Additional Labor to Manage Humanlike GenAI} 
Using and managing humanlike genAI introduced additional labor burdens. Beyond typical new technology adjustments, workers faced disrupted workflows and the added burden of making professional decisions without adequate guidance. Workers specifically highlighted missing frameworks to guide them as users, designers, evaluators, and communicators about humanlike systems. For engineers, this involved challenges translating abstract humanlikeness into technical features. One engineer (P19) working on an AI system for interactive data dashboards expressed uncertainty about controlling humanlike characteristics, stating ``it is harder to create guidelines for a humanlike AI because, how can you even define what it is to be human?'' Other engineers echoed this sentiment. While some high-level guidance exists, specific guidelines felt like an elusive task. P22, a ML engineer at a digital infrastructure consulting firm explained, ``as we’re applying it more to our respective work, I think it gets harder to have more descriptive guidelines.''

Existing guidance felt obsolete when applied to humanlike genAI. With an eye toward fraud, policy workers emphasized the hazards of outdated guidance in identifying scammers, such as recommending users look for typos or delayed responses, no longer apply to sophisticated AI.  P24, a policy analyst at a large software company noted: ``I don't know if there actually is any traditional wisdom that we can really give these people anymore.''
Missing and out-of-date guidance ultimately increased workload as workers navigate new systems while facing maintained or increased performance expectations. Added pressure was clear for one policy analyst in privacy and security at a gaming company, who found it increasingly hard to tell true sensitive documents from AI-generated ones.

\begin{quote}
I … uh … actually left my last position because [humanlike genAI] was … in full honesty, creating so much of an issue so quickly that … ummm … the company that I was with was not keeping up with it. (P26)
\end{quote}
Here, ``humanlike'' refers to generated \textit{outputs}  carrying the fidelity of human-authored or official documents. Fraudsters increasingly used humanlike genAI to falsify realistic identification documents at her prior transportation tech startup, leaving her unable to manage the growing fraud volume effectively and without erroneously flagging legitimate documents.

Workload concerns posed  challenges for small or under-staffed teams. A communications worker at a small company of less than 50 people (P8) described how they were grateful to work with a large quality control team, highlighting ``not everybody has the ability and the time to check things.'' Without adequate support, the risk of inaccuracies and errors increases significantly.

\subsubsection{The Hazard of Miscalibrated Trust} 
Humanlike features, particularly high-fidelity outputs, contribute to miscalibrated trust in AI systems. Tech workers expressed significant concern that humanlikeness  fosters a false sense of reliability and trust shaping their concerns as both users and developers, linking to fluid, natural language and tone, which can obscure errors. P7 (Content Manager) noted:
\begin{quote}
[Humanlikeness] makes it very intuitive to interact with because ...  it just answers like a person but it also does make it harder to detect errors because you’re like ‘okay that’s a good answer and if you’re not careful you just run with it.
\end{quote}

Engineers similarly emphasized a gap between others' trust of humanlike systems and their own hesitations rooted in experiences evaluating model errors, with P21 adding, ``I still use Stack Overflow and Stack Exchange for 99\% of my error tracking. Why? Because I trust it.''

Beyond individual outputs, workers worried that humanlikeness might lead users to misinterpret AI capabilities, leading to  over-reliance and insufficient critical evaluation.

This overtrust was feared to be pronounced among certain user groups, such as children. Although humanlikeness may influence trust independent of the psychosocial process of anthropomorphism, its theory predicts higher anthropomorphism among children due to relatively less exposure to non-humanlike agents \cite{epley2007seeing}. One engineer (P19) emphasized the need for better AI training and responsible communication rather than blaming users for interpreting it a certain way: ``I think our responsibility there is to better train the AI and not obviously blame the children for misinterpretations.''
This intuition reflects that anthropomorphism-- and sensemaking--stems from a human tendency to  infer mental states based on their own experiences \cite{keysar2002self, nickerson1999we}.
However, workers grappled with conveying system limitations effectively, recognizing the need to support users in understanding AI decision-making even in non-public-facing technologies: 
\begin{quote}
[Our technologies] don’t have public-facing interactions. That is a blessing in disguise. But, you still have to ensure that when individuals are faced with the veracity of the information, that you provide them support to understand how the system got to its conclusion. (P12, UX)
\end{quote}

\subsubsection{The Hazard of Believability} 
Believable content refers to AI-generated material convincingly realistic enough to be mistaken for human-generated content. While workers across roles acknowledged the potential,  policy workers, being especially attuned to fraud and abuse, expressed incisive concerns. According to P24 (Policy Analyst), humanlike genAI is already a ``thorn in their side'' enabling bad actors. They anticipated believable content would increase scam success rates, increasing end-user vulnerability while simultaneously  challenging workers to proactively detect malicious content. P25, a policy analyst at a customer service startup further explained: ``...detecting it on our side, and stopping it if it looks credible [is harder]...we’re going to have a harder time figuring out what’s bad and stopping it''

For policy workers, believable content exacerbated role challenges, leading to additional labor for them as non-user stakeholders and a constant struggle against evolving threats. The rapid pace of AI updates necessitated continuous adaptation, making proactive strategies ``hard to get in front of'' (P24). P26, the policy worker overwhelmed by AI-generated identification documents, described constantly playing catch-up: ``I was always kind of playing reactive ... trying to become more proactive, trying to get ahead of it. But, in full honesty, I was never able to.''

\subsubsection{The Hazards of Worsened User Experience and Social Interaction} UX workers and technical writers shared concerns about humanlike genAI's impact on user experience, identifying  hazards rooted in diminished quality compared to human interaction. One startup UX researcher (P11) highlighted inconsistent system transparency and explainability: ``...Am I chatting with a human or a robot? Do I have to change how I am talking [to be understood]? Am I feeding into the all-knowing thing?'' She also recounted an anecdote where a participant in a remote study questioned whether she was an AI, indicating that increasing believability of AI-generated content-- outside of what she uses or develops-- challenges rapport-building necessary for interview studies:

\begin{quote}
What has made [my job] harder is the prevalence of [humanlike genAI]. Like you, I’m a researcher and I have to get people to open up… and now I have to prove that I am a person and get you to talk to me about things.
\end{quote}

Other UX workers and technical writers echoed that humanlike genAI hindered positive user and consumer relationships, even when AI content was imperfect. A content manager for an online sales company (P7) emphasized the importance of establishing credibility and trust in promotional content, noting AI factual errors or bizarre hallucinations risk undermining brand authority and ``just looking dumb.’’  Some argued relying on humanlike genAI, regardless of output quality, inherently jeopardizes brand reputation. As one self-employed UX worker (P13),  put it: ``At some point if you don’t have any humans working for you …you aren’t going to be seen as an authentic company.’’

Finally, workers worried human-AI interaction might supplant genuine human connection, weakening human interaction long-term. While less tied to professional experience, this reflected broader anxiety about developing humanlike genAI products, echoing extant ethical concerns~\cite{gabriel2024ethics}. One product manager (P14) reflected: ``as more and more technology becomes humanlike, I worry the next generation is going to conflate interactions with AI as the same as interactions with other human beings. And how solitary that may be for them and for this society, in general.''

\subsubsection{Disparate AI Literacy} A cross-cutting hazard was users' varying technical knowledge. All workers emphasized  humanlikeness could mislead users  with low AI literacy \textit{and} workers managing these systems, potentially leading to over-estimations of system capabilities:

\begin{quote}
The humanlike nature of it itself...could be more misleading to a fault. Where they may get the impression that what they're interacting with is… is much smarter or more intuitive than it actually is. (P20, ML engineer)
\end{quote}
A prevalent concern involved the increased AI knowledge needed to recognize potential system abuse. As P24, a policy worker, stated: ``it's going to widen the pool of people who are vulnerable to this sort of thing.'' Indeed, even workers with greater experience with humanlike genAI acknowledged their own knowledge gaps. P29, a technical writer for a human resources software company, lamented learning about risks from engineers about her own use of genAI: ``there's a lot of issues out there that I wouldn't even think about watching for...''

\subsubsection{The Hazard of Worker Replacement} 
Communications and PR workers were especially concerned about job replacement by humanlike genAI. While not unique worker concerns \cite{woodruff2024}, humanlike genAI might be particularly threatening to creative skills at the heart of their work. For instance, P8 a freelance tech writer who was hesitant to use genAI in her work, acknowledged the potential benefits of humanlike genAI but at the same time voiced concerns about its impact on skill development, ``So if students start using it to write papers and things like that, they'll never learn,'' she observed.
Beyond students, she further wondered if ``people are just not going to learn how to be creative anymore'' as AI increasingly generates content and ideas for them.
 
Workers also feared broader homogenization of creative outputs as AI-generated content becomes more prevalent. P6, in marketing, reflected on her experience using chatbots to mimic writing styles, ``but the real problem to me is that if it becomes an echo system where it just becomes everything that’s out there on the Internet was written by ChatGPT, then it starts to become more and more sameness.'' This observation highlights the potential for humanlike genAI to stifle originality and promote bland uniformity in creative expression.

\subsection{Filling Guidance Gaps with Positional Knowledge}
\label{positional-differences}

A distinct shared concern was the lack of guidance surrounding appropriate genAI use despite increasing pressures to adopt it. This lack of standardized guidance forced workers to draw on a variety of individual experiences to form judgments. Reflecting this, we observed distinct patterns in the hazards most salient to each job role.

\subsubsection{Mounting Pressure with Little Guidance}
Workers agreed on clear interest in the field and their workplaces to adopt genAI due to ``AI fever'' (P21). This excitement created pressure to adopt, use, or integrate the technology into their workflows, fueling uncertainty and skepticism regarding the personal relevance of genAI. For example, one UX worker responded to top-down pressures from job superiors citing her company's  research finding users were reticent toward humanlike genAI features:
\begin{quote}
[It wouldn't make sense] if we just went off of assumptions based on higher ups' opinions because there wasn’t really a thirst for humanlikeness [from users] on the output side versus the input side. (P9, UX)
\end{quote}
Still, workers agreed that learning how to use genAI was becoming a necessary job skill. P28, a technical writer shared, ``One of the other qualifications is you need to be able to fact check what the AI told you, so we can make sure that our AI is trustworthy’’, and a product manager more sharply stated, ``...it's evolve or die. You're gonna have to learn to use these automations to do more of your work’’ (P15, PM).

\label{lowinformation}
Despite this urgency to use genAI, workers reported having little usable guidance by way of best practices or company policies. Workers emphasized the need for informed guidance, citing insufficient company policies, lack of role-specific best practices, and missing field-wide standards. P1, a content strategist, noted there was ``no oversight'' or ``thought behind the use,'' concluding that ``there's no shepherd that's guiding'' the work. 
She  described a chaotic onboarding experience: ``…when I started my new job recently ... people had been using ChatGPT to write stuff for our website … and there were no guidelines at all,''  raising concerns about  product quality issues, including accidental plagiarism, potentially arising from a lack of protocols outlining genAI use and content review.

Workers underscored the urgency of developing internal policy and  principled approaches to minimize undesired consequences, such as clarifying how genAI systems work and dis/allowed data practices: 
\begin{quote}
I’ve been trying to get my company to come up with something. One of the dangers is that I don’t want [Open AI] to train on my campaigns because I think they are pretty good. So by giving ChatGPT my data they could use it right? …we have no policy whatsoever. (P6, PR)
\end{quote}
However, keeping policies up-to-date was a nontrivial concern, as workers flagged the rapid pace of genAI development. P26 (policy), who worked on developing internal policies for genAI use, noted, “I found myself re-writing processes and changing things almost weekly, sometimes, very, very quickly... it was a lot,” while, P20, an ML engineer at a small B2B startup, explained inherent technical challenges to creating up-to-date policies “...you might be on model version one [then] you switch to model version two, and then all of a sudden your guidelines ... for version one no longer apply.”  The current development landscape thus challenges the maintenance of clear guidance and responsible use.
Ultimately, limited-- and often absent-- guidance exacerbated humanlikeness hazards  and prevented principled  development  for system development and use.

\subsubsection{Implicit Standpoints}
\label{standpoints}

Given the lack of formal guidance, workers relied on their own varied perspectives as both users and non-users. While qualitative differences in  insights related to user vs. non-user perspectives were not observed,  the standpoints from which workers  sensemade about hazards was fluid. For example, the hazards of `worsened UX and social interaction' and `disparate AI literacy' were discussed from workers' standpoints as users of genAI, whereas the hazard of `additional labor to manage humanlike genAI' was discussed from workers' perspectives as non-user stakeholders (i.e., a hazard as a result of \textit{others'} use of genAI). Meanwhile, `miscalibrated trust' combined standpoints regarding workers' mistrust as users and their perceived responsibility to improve system accuracy as product and model developers.

While workers across  roles shared perspectives on the nature of humanlike genAI, their attention to hazards differed, reflecting situated expertise. For example, of the 35 quotes in our coded data that explicitly discussed hazards leading to abuse and fraud, 27 came from the policy workers. Meanwhile, communication workers and content writers, who frequently craft public written material, discussed genAI hallucinations and factual errors more than any other hazard, highlighting the potential for plagiarism and erosion in brand trust. For example, P7, a content manager, discussed an example of risks stemming from subtle errors in AI-generated suggestions for brand blog copy:

\begin{quote}
...[for example] suggesting an article on how to clean the filter in your microwave, which, if you don't think very hard about it, you're like, `okay sure,' but you can't clean those; you just have to throw them away... So it might seem like a small error, but if somebody's reading our blog and they're like, `None of your tips make sense,' that damages [brand] authority. (P7, Content Manager)
\end{quote}
In contrast, ML engineers, who frequently assess factuality, focused on humanlikeness as a component of AI model quality rather than as a mechanism for downstream threats to brand trust, like communication workers did, or as a mechanism for potential fraud, like policy workers did. One engineer related a specific example of incorrectly generated Linux code, emphasizing a preference for accuracy before humanlikeness,
\begin{quote}
...if I would have executed that command, it would have been horribly dangerous to the system. And so to me, the accuracy is so much more important than the humanlike aspect of it. And unfortunately, I think the human aspect of it is misleading. (P20, ML Engineer) 
\end{quote}
Past job role experience also contributed to different  salient hazards. Policy workers, for instance, recalled how the emergence of easy-to-use \textit{script kiddies} and phone-based PDF file editors led to new fraud types. P26 observed that ``everybody has a phone in their pocket nowadays and it's really easy for anybody to make anything, at any time [with genAI].''
These past experiences with fraud informed their expectations of a monotonic relationship between increased user-friendliness tied to humanlikeness and new forms of fraud detection challenges. Thus, both on-the-ground genAI experiences and past job experiences formed important sources for  grounded assessments.
\section{Discussion: Navigating Unsettled Terrain}
\label{discussion}
Our discussions with tech workers indicated numerous hazards rooted in humanlike genAI. Addressing these hazards is complicated by (1) the intersecting, differing, and often competing ways in which ``humanlikeness'' in AI is conceptualized and (2) prevailing knowledge gaps amid pressures to adopt genAI rapidly. Effectively navigating this ambiguity requires shared, coherent knowledge about genAI systems, plus structured guidance and resources to cultivate this understanding further. Our findings suggest conceptual and practical organizational pathways toward responsible development, encompassing policies, practices, and improved guidance.

\subsection{Toward Developing Coherent Knowledge on ``Humanlikeness'' in GenAI}

Holistic sociotechnical evaluation of ``humanlike genAI'' requires understanding the interplay of user perceptions and humanlike design features. Assessing humanlikeness is challenging because, as tech workers indicated, humanlike qualities stem from diverse sources (e.g., underlying model capabilities, interface design, inherent task sophistication). This contrasts with the AI field's heavy reliance on relatively narrow benchmark evaluations. Currently, few benchmarks exist for assessing humanlikeness directly, and due to diverse and often imprecise definitions, humanlikeness is not easily quantifiable for systematic evaluation.

\begin{figure*}
\begin{subfigure}[b]{0.9\textwidth}   
        \centering
        \caption{Reductive view of anthropomorphism}
        \includegraphics[width=\textwidth]{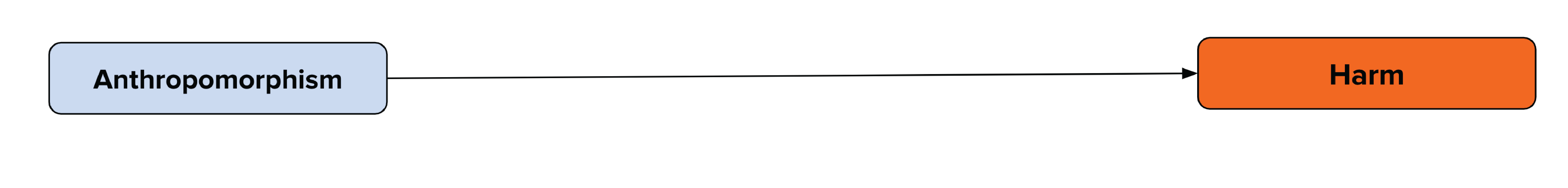}
        \label{fig:first}
        \Description[Flow chart labeled ``(a) Reductive view of anthropomorphism'' connected by flow links.]{The start state is ``Anthropomorphism'' which flows to the end state “Harm”}
    \end{subfigure}
    \begin{subfigure}[b]{0.9\textwidth}
        \centering
        \caption{Re-conceptualized view of anthropomorphism}
        \includegraphics[width=\textwidth]{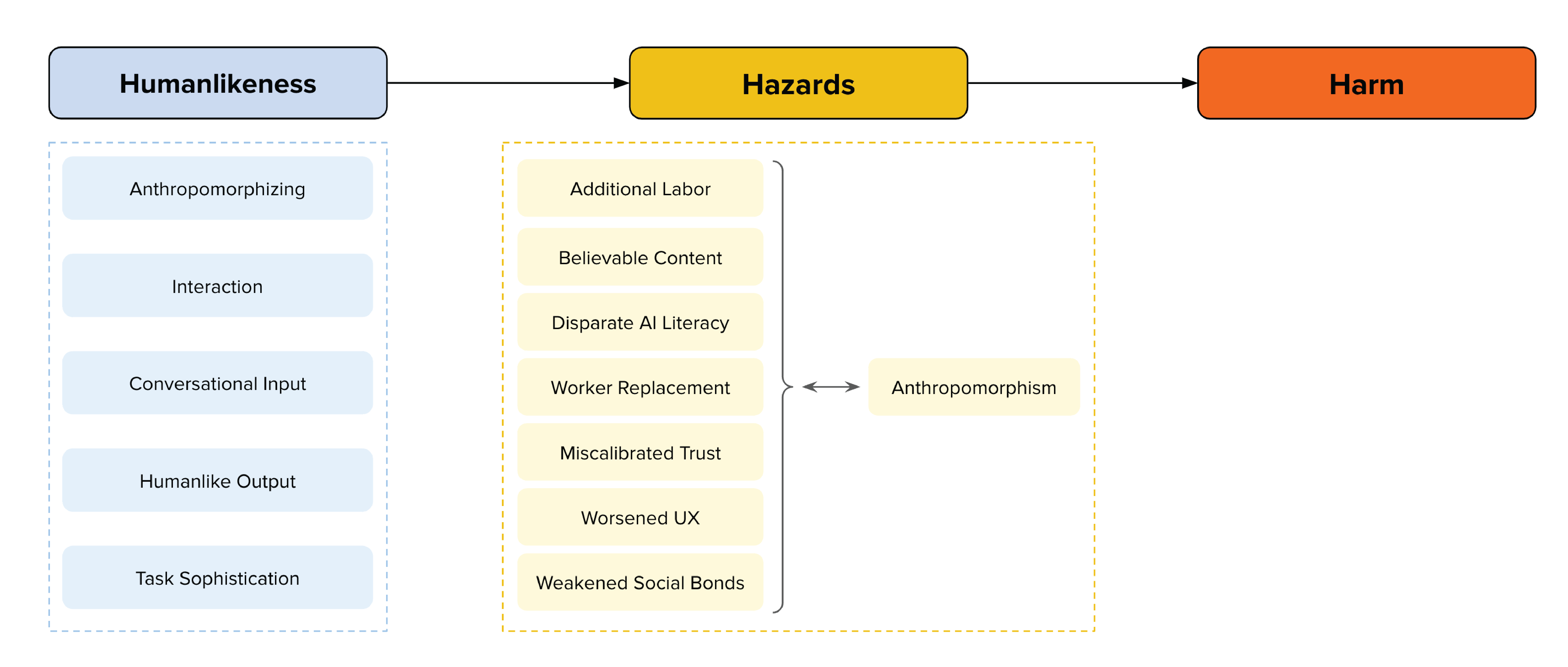} 
        \label{fig:second}
        \Description[Flow chart labeled ``(b) Re-conceptualized view of anthropomorphism'' connected by flow links.]{The start state is ``Humanlikeness'' which flows to ``Hazards''. ``Hazards'' flows to the end state “Harm”. ``Humanlikeness'' contains the sub-categories Anthropomorphizing, Interaction, Conversational Input, Humanlike Output, Task Sophistication. ``Hazards'' contains the sub-categories Additional Labor, Believable Content, Disparate AI Literacy, Worker Replacement, Miscalibrated Trust, Worsened UX, Weakened Social Bonds, and Anthropomorphism. A double-headed arrow Anthropomorphism and the other hazards indicates that Anthropomorphism can be a cause of, caused by, or mutually reinforce them.}
    \end{subfigure}
    \caption{Two approaches to conceptualizing humanlikeness/harm relationships. (a) depicts a reductive view of anthropomorphism which 1) conflates humanlikeness with anthropomorphism and 2) posits that anthropomorphism necessarily leads to harm. (b) depicts our re-conceptualization. This view posits that humanlikeness produces hazards which have the \textit{potential} to produce harm-- with Anthropomorphism among them and interacting with other hazards. Humanlikeness in this view consists in varied categories (described in \ref{humanlikeness}), and the hazards they produce are similarly varied (described in \ref{hazards}).}
\label{fig:concept-map}
\end{figure*}

In this light, our study reveals implicit relationships between humanlike genAI features, anthropomorphism, and other related hazards. Figure \ref{fig:concept-map} illustrates two contrasting conceptions of the relationship between humanlikeness, hazards, and harms. Figure \ref{fig:first} depicts a common, yet flawed conceptualization in which anthropomorphism and humanlikeness are conflated. However, as discussed in \ref{humanlikeness-in-genai}, anthropomorphism constitutes a condition with the \textit{potential} for harm (i.e., a hazard), rather than a harm in itself. Figure \ref{fig:second} combines this understanding of anthropomorphism as a \textit{hazard} with insights from tech worker perspectives gathered in our study to highlight other hazards that sit alongside it. Thus, it distinguishes anthropomorphism from the humanlike features that may facilitate it or dynamically interact with it. This updated mapping unpacks ``humanlikeness'' terminology and provides a clearer conceptual basis for studying and measuring the impacts of humanlike features across modalities and interfaces. 


In our reconceptualization of anthropomorphism as a hazard among other hazards, we do not fundamentally shift our understanding of the theory. Instead we bring clarity to its implications for technology development and use. Our approach focuses on clarifying the conceptual relations between anthropomorphism, contributing factors (i.e., humanlike features), and \textit{other} potential hazards. This enables investigation into how humanlike design may be calibrated to facilitate or mitigate anthropomorphism as deemed necessary. Our clarification provides two primary benefits:

\textbf{Specificity in Research:} Specifying categories of humanlikeness in genAI allows for more nuanced research into how these features do or do not facilitate hazards, broadening humanlikeness safety discussions beyond anthropomorphism. Even if a humanlike features does not exacerbate or lead to anthropomorphism, it may still introduce other hazards. Thus, humanlikeness remains a concern even in the absence of anthropomorphism. Moreover, the direction of the causal relationship between specific hazards and anthropomorphism may vary. For instance, anthropomorphism can function as both a cause (e.g., introducing increased labor) and an effect  (e.g., be influenced by a user's AI literacy). Finally, some relationships, such as those between believable content and anthropomorphism, might be mutually reinforcing. 

\textbf{Targeted Mitigation:} Disaggregating potential relationships between humanlikeness categories and various hazard categories clarifies areas for assessment, measurement validation, and mitigation. In this way, categories of humanlikeness can be applied to systems that cut across modalities and interfaces to scaffold empirical studies of specific features. For example, iconography in interactive chatbot dialogue interfaces can be studied in relation to trust calibration, while pitch and prosody in humanlike output can be studied in voiced systems. This is a critical conceptual distinction as strategies cannot be applied directly to anthropomorphism itself-- due to its psychosocial nature-- but rather must be applied to specific design elements, model capabilities, or usage policies. Key measurement questions include identifying which humanlikeness features associate most strongly with specific hazards and determining which metrics can appropriately evaluate hazard frequency and severity.

Ultimately, our conceptual map provides a clearer framework for validating the effects of humanlikeness and anthropomorphism (and other hazards) in genAI systems. Further research into their impacts is key for informing development strategies. If empirical investigation reveals certain humanlike features do not significantly influence anthropomorphism or contribute to hazards (or benefits) in a given context, they can be de-prioritized. Prioritization can draw upon safety engineering  principles~\cite{rismani2023, dobbe2024toward, leveson2016engineering}, such as assessing risk as a function of hazard likelihood and severity. Importantly, hazard assessment  should employ mixed methodologies and incorporate meaningful stakeholder engagement to consider risks beyond those directly affecting the firms producing humanlike genAI systems.

\subsection{Workers Need Support in the Face of Critical Knowledge Gaps}
Our tech workers' experiences speak to ongoing challenges facing tech workers more broadly, as well as  challenges soon to face users and workers in other professional domains. With limited guidance and visibility into the upstream development of foundation models conducted at major AI labs, workers rely on professional training, firsthand interactions with AI, and work-related goals to partially fill knowledge gaps. That is to say, workers have expertise on genAI developed not only through professional training, but also through formal and informal genAI use, as well as through interaction with others' use. We found that workers implicitly wear these simultaneous and constantly shifting ``hats'' while sensemaking, yet their collective judgments illuminate a broad hazard landscape and the decision support needed to navigate it in professional domains. This reliance on individual interpretation and experience, however, can lead to inconsistencies and ad hoc approaches in navigating hazards of humanlike AI.

\subsubsection{Tech Workers Sensemake to Fill Knowledge Gaps}
In the absence of guidance, workers are forced to independently assess humanlike genAI systems, relying on interactions across personal and professional spheres. This sensemaking involved weighing costs to their performance, well-being, and ultimately, accountability for the AI's output. For example, concerns about reputation and livelihood directly informed choices to avoid ``beta tech’’ in their professional lives.
The different hazards workers raised reflect \textbf{professional positionality}, which plays a key factor in hazard reasoning. Aligning with work by \citet{scheuerman2024products} on positionality in computer vision design, our study shows how tech workers leverage positional knowledge to identify salient genAI hazards. However, worker reasoning also revealed \textbf{positional knowledge gaps}, related to their roles. To some degree, knowledge differences about hazards is expected because they speak to their expertise and salient issues in specific job domains. However, in AI, engineers are primarily in charge of crafting and implementing AI evaluations, meaning hazards salient to workers in other roles are not necessarily salient to those in charge of evaluation. From a RAI perspective, the distributed nature of AI development means worker positionality informs disparate styles of thinking about humanlikeness that may hinder coordinated mitigation. Worker positionality, thus, influences both knowledge creation and application.

Workers primarily spoke from perspectives rooted in job function, rather than their company or organization positions on how humanlike AI should be considered. As genAI use expands to more fields and organizations, workers may similarly exhibit different positional knowledge and attention toward salient issues, producing a varied array of knowledge gaps according to both individual role and industry. Our work highlights the importance of attending to specific worker roles, rather than homogenizing knowledge within organizations, as roles surface different insights about potential hazards and harms. In other words, we must think about controls and mitigations from a role-oriented perspective as these elicit different aspects of harm.

\subsubsection{Workers Need Holistic and Cross-Functional Support}
Workers' AI encounters and hazard reasoning reveal a critical need for support and point to widespread information deficits for working responsibly with humanlike genAI. Despite pressure to adopt AI, workers expressed a  cautious, ambivalent perspective, stemming from pragmatic awareness of the potential material consequences of adopting poorly understood technology, as well as limited capacity to reflect on genAI's usefulness or hazards amidst work demands. This echoes broader challenges experienced by RAI practitioners, who pursue responsible development while navigating organizational structures with limited guidance~\cite{rakova2021responsible}. While the tech workers in our study were not RAI practitioners by job mandate, they expressed desires to avoid the hazards they discussed, and the issues they raised point to \textit{organizational} RAI challenges.  Workers' desire for contextualized support regarding ``humanlike'' genAI echoes struggles of ``non-RAI'' tech workers adapting documentation frameworks~\cite{heger2022understanding} or addressing root causes of fairness hazards~\cite{holstein2019improving}.  Our conceptual mapping helps clarify facets of ``humanlikeness'' requiring further specification as prerequisite for generating use case-specific guidance.

A major RAI challenge is the distributed nature of genAI development, particularly the separation between foundation model development and product development across different organizations. This institutional barrier limits foundation model transparency and hinders information flow between upstream and downstream development pipelines, exacerbating knowledge gaps arising from the novelty of ``humanlike'' genAI.
\citet{rakova2021responsible} describe organizations as co-existing within ecologies of formal and informal coordination, with relationships between organizations oriented toward various goals, including standards and best practice development and staying abreast of industry trends. Strong coordination within and across organizations is critical to address ``humanlikeness hazards,'' particularly as downstream organizations and users struggle with the rapid pace of change.  Even with appropriate development practices, unexpected interactions between product features or components can cause unanticipated harms~\cite{slota2023many}. Therefore, ``humanlikeness'' mitigations must be distributed throughout development and use decisions, responsive to issues identified within different tech workers' expertise domains. Downstream workers  need information to understand foundation model constraints and to evaluate ``humanlike'' genAI systems contextually. Conversely, upstream developers need downstream insights to ensure effective and appropriate guardrails at the foundation model level. 

While our conceptual map integrates diverse perspectives, not all organizations possess such a range of internal expertise. For smaller organizations using genAI in workflows or building upon third-party foundation models, our hazard map offers broader insights than might be  internally available. Still, a broader need remains to map how ``humanlikeness'' hazards emerge across different development stages and use contexts. Lack of end-to-end understanding can lead to disparate assumptions about ``humanlike'' features and their impacts, hindering effective mitigation and risking user harms.

\subsection{``Humanlikeness'' Requires Specificity}
Supporting users in identifying humanlikeness hazards and improving coordination around addressing them requires deeper clarity on the different forms humanlikeness can take. The unsettled knowledge environment obscures what ``humanlikeness`` does and can mean, which is a hurdle to developing effective mitigations.
The diverse ways workers perceive ``humanlikeness'' mirrors the multiple conceptualizations of ``anthropomorphism'' in AI research: as a system feature, interaction experience, and user tendency~\cite{li2022anthropomorphism}. AI research sometimes uses ``anthropomorphism'' interchangeably with ``humanlikeness,'' even without measuring the psychosocial process (e.g., \cite{de2016almost, ha2021exploring, troshani2021we}). Thus, ``humanlikeness’’ and ``anthropomorphism'' function as \textit{suitcase words}-- combining and conflating disparate meanings, but referencing  common aspirations~\cite{lipton2019,minsky2007emotion}, such as developing  AI systems that match or extend human capabilities, which~\citet{lipton2019} emphasize as a unifying goal across AI. 
Consequences of \textit{suitcase words} include misaligned goals  and hindered  application of abstract guidelines~\cite{rakova2021responsible}. As \citet[p. 3]{rakova2021responsible} highlight, disparate understandings of core terminology can misdirect attention to ``future, similarly abstract technologies rather than current, already pressing problems.’’ Ambiguous terminology is a barrier  to generating even abstract guidance for development and use. Consequently, humanlikeness as described in advertising, policies, or interface design strategies, can reference vastly different concepts.
Moreover, unpacking humanlikeness is challenging due to its historical absence as an RAI focus  and the difficulty of  identifying aspects of humanlikeness masked by frictionless~\cite{masrani2023, cox2016} UI design. These issues compound RAI coordination challenges \citet{rakova2021responsible} discussed in the context of AI integration into production-ready systems, extending to non-tech stakeholders whose work is transformed by their own or others' use of humanlike genAI. 

\subsubsection{Specificity for Developers}

Ambiguity of ``humanlikeness'' and ``anthropomorphism'' necessitates clearer operationalization to support harm-reductive AI development. Although most tech workers in our study were not direct model developers, their insights underscore that clear definitions are essential for assessing system design and behavior suitability for downstream contexts. This need is heightened by the tendency within ML development to characterize system behavior post-hoc~\cite{hurshman2024opaque}. Claims of ``humanlikeness'' often follow this norm, leading to a dearth of standardized assessments to empirically characterize its presence and effects.

In response, ``humanlikeness'' should be clearly articulated in product development goals. Specificity may reveal its implicit nature in many genAI capabilities. Indeed, it may be rooted in the core function of a system, as was the case for the product manager who described her Tesla as humanlike by virtue of its self-driving capabilities. Thus, reducing humanlikeness, in some cases, may diminish the core function of a system. Moreover, the same features can elicit different interactions and judgments of humanlikeness from different users~\cite{ljungblad2012hospital}. Rather than broadly aiming to reduce humanlikeness, developers should identify specific humanlike features and determine measurement thresholds where hazards decrease meaningfully while preserving desired user experiences or product outcomes. Tech workers affirmed that humanlike features and anthropomorphism are not inherently negative. Thus, developers must control humanlike features to mitigate harms, but cannot simply adopt a reductive stance to eliminate ``humanlikeness'' solely to reduce anthropomorphism.

While ``humanlikeness’’ can indicate a general system style, lack of specificity obscures salient aspects for given use cases. Our work demonstrates that optimizing design for a given humanlikeness heuristic   may inadvertently lead to undesirable product features and affordances. For example, designing for humanlike outputs is distinct from designing for humanlike interaction dynamics. More specifically, optimizing for rapid, humanlike ad copy generation can make the system subpar as a writing collaborator, which requires sensitivity to the cadence of interaction dictated by the human user. Benefits like accessibility and productivity exist, as workers noted, regarding natural language capabilities (see also: specific uses in disability art communities~\cite{bennett2024}). However, evaluating specific aspects of ``humanlikeness'' is requisite for exploring potential affordances contextually.

\section{Conclusion}
Anthropomorphism is not new, but genAI systems mark a clear progression toward systems increasingly indistinguishable from humans in various ways. This leaves workers in an unsettled knowledge environment regarding the integration of these technologies into their workflows and end-user services. Researchers have begun critically examining humanlike AI, yet empirical work remains scant. Through focus groups, we surfaced  challenges workers face coordinating responsible use, a range of hazards posed by humanlike genAI, and varying conceptions of ``humanlikeness.'' Drawing on workers' situated expertise, we contribute a conceptual map articulating connections between identified hazards and varying conceptions of ``humanlikeness.'' This map charts a path towards supporting researchers' and practitioners' coherent understanding of humanlikeness and toward addressing the complex relationship between humanlike features, perceived hazards, and RAI development.

\bibliographystyle{ACM-Reference-Format}
\bibliography{sample-base}


\begin{thebibliography}{156}


\ifx \showCODEN    \undefined \def \showCODEN     #1{\unskip}     \fi
\ifx \showISBNx    \undefined \def \showISBNx     #1{\unskip}     \fi
\ifx \showISBNxiii \undefined \def \showISBNxiii  #1{\unskip}     \fi
\ifx \showISSN     \undefined \def \showISSN      #1{\unskip}     \fi
\ifx \showLCCN     \undefined \def \showLCCN      #1{\unskip}     \fi
\ifx \shownote     \undefined \def \shownote      #1{#1}          \fi
\ifx \showarticletitle \undefined \def \showarticletitle #1{#1}   \fi
\ifx \showURL      \undefined \def \showURL       {\relax}        \fi
\providecommand\bibfield[2]{#2}
\providecommand\bibinfo[2]{#2}
\providecommand\natexlab[1]{#1}
\providecommand\showeprint[2][]{arXiv:#2}

\bibitem[Abercrombie et~al\mbox{.}(2021)]%
        {abercrombie-etal-2021-alexa}
\bibfield{author}{\bibinfo{person}{Gavin Abercrombie}, \bibinfo{person}{Amanda Cercas~Curry}, \bibinfo{person}{Mugdha Pandya}, {and} \bibinfo{person}{Verena Rieser}.} \bibinfo{year}{2021}\natexlab{}.
\newblock \showarticletitle{{A}lexa, {G}oogle, {S}iri: What are Your Pronouns? Gender and Anthropomorphism in the Design and Perception of Conversational Assistants}. In \bibinfo{booktitle}{\emph{Proceedings of the 3rd Workshop on Gender Bias in Natural Language Processing}}, \bibfield{editor}{\bibinfo{person}{Marta Costa-jussa}, \bibinfo{person}{Hila Gonen}, \bibinfo{person}{Christian Hardmeier}, {and} \bibinfo{person}{Kellie Webster}} (Eds.). \bibinfo{publisher}{Association for Computational Linguistics}, \bibinfo{address}{Online}, \bibinfo{pages}{24--33}.
\newblock
\href{https://doi.org/10.18653/v1/2021.gebnlp-1.4}{doi:\nolinkurl{10.18653/v1/2021.gebnlp-1.4}}


\bibitem[Abercrombie et~al\mbox{.}(2023)]%
        {abercrombie2023mirages}
\bibfield{author}{\bibinfo{person}{Gavin Abercrombie}, \bibinfo{person}{Amanda~Cercas Curry}, \bibinfo{person}{Tanvi Dinkar}, \bibinfo{person}{Verena Rieser}, {and} \bibinfo{person}{Zeerak Talat}.} \bibinfo{year}{2023}\natexlab{}.
\newblock \bibinfo{title}{Mirages: On Anthropomorphism in Dialogue Systems}.
\newblock
\showeprint[arxiv]{2305.09800}~[cs.CL]
\urldef\tempurl%
\url{https://arxiv.org/abs/2305.09800}
\showURL{%
\tempurl}


\bibitem[Adam et~al\mbox{.}(2021)]%
        {adam2021ai}
\bibfield{author}{\bibinfo{person}{Martin Adam}, \bibinfo{person}{Michael Wessel}, {and} \bibinfo{person}{Alexander Benlian}.} \bibinfo{year}{2021}\natexlab{}.
\newblock \showarticletitle{AI-based chatbots in customer service and their effects on user compliance}.
\newblock \bibinfo{journal}{\emph{Electronic Markets}} \bibinfo{volume}{31}, \bibinfo{number}{2} (\bibinfo{year}{2021}), \bibinfo{pages}{427--445}.
\newblock


\bibitem[Akbulut et~al\mbox{.}(2024)]%
        {akbulut2024all}
\bibfield{author}{\bibinfo{person}{Canfer Akbulut}, \bibinfo{person}{Laura Weidinger}, \bibinfo{person}{Arianna Manzini}, \bibinfo{person}{Iason Gabriel}, {and} \bibinfo{person}{Verena Rieser}.} \bibinfo{year}{2024}\natexlab{}.
\newblock \showarticletitle{All Too Human? Mapping and Mitigating the Risk from Anthropomorphic AI}.
\newblock \bibinfo{journal}{\emph{Proceedings of the AAAI/ACM Conference on AI, Ethics, and Society}} \bibinfo{volume}{7}, \bibinfo{number}{1} (\bibinfo{date}{Oct.} \bibinfo{year}{2024}), \bibinfo{pages}{13--26}.
\newblock
\href{https://doi.org/10.1609/aies.v7i1.31613}{doi:\nolinkurl{10.1609/aies.v7i1.31613}}


\bibitem[Amershi et~al\mbox{.}(2019)]%
        {amershi2019}
\bibfield{author}{\bibinfo{person}{Saleema Amershi}, \bibinfo{person}{Dan Weld}, \bibinfo{person}{Mihaela Vorvoreanu}, \bibinfo{person}{Adam Fourney}, \bibinfo{person}{Besmira Nushi}, \bibinfo{person}{Penny Collisson}, \bibinfo{person}{Jina Suh}, \bibinfo{person}{Shamsi Iqbal}, \bibinfo{person}{Paul~N. Bennett}, \bibinfo{person}{Kori Inkpen}, \bibinfo{person}{Jaime Teevan}, \bibinfo{person}{Ruth Kikin-Gil}, {and} \bibinfo{person}{Eric Horvitz}.} \bibinfo{year}{2019}\natexlab{}.
\newblock \showarticletitle{Guidelines for Human-AI Interaction}. In \bibinfo{booktitle}{\emph{Proceedings of the 2019 CHI Conference on Human Factors in Computing Systems}} (Glasgow, Scotland Uk) \emph{(\bibinfo{series}{CHI '19})}. \bibinfo{publisher}{Association for Computing Machinery}, \bibinfo{address}{New York, NY, USA}, \bibinfo{pages}{1–13}.
\newblock
\showISBNx{9781450359702}
\href{https://doi.org/10.1145/3290605.3300233}{doi:\nolinkurl{10.1145/3290605.3300233}}


\bibitem[Anthropic(2024)]%
        {anthropic2024}
\bibfield{author}{\bibinfo{person}{Anthropic}.} \bibinfo{year}{2024}\natexlab{}.
\newblock \bibinfo{title}{Release Notes: System Prompts}.
\newblock
\urldef\tempurl%
\url{https://docs.anthropic.com/en/release-notes/system-prompts}
\showURL{%
\tempurl}


\bibitem[Apps(2024)]%
        {miAI2024}
\bibfield{author}{\bibinfo{person}{MiAI Apps}.} \bibinfo{year}{2024}\natexlab{}.
\newblock \bibinfo{title}{MiAI}.
\newblock
\urldef\tempurl%
\url{https://miaiapps.com/}
\showURL{%
\tempurl}


\bibitem[Araujo(2018)]%
        {ARAUJO2018183}
\bibfield{author}{\bibinfo{person}{Theo Araujo}.} \bibinfo{year}{2018}\natexlab{}.
\newblock \showarticletitle{Living up to the chatbot hype: The influence of anthropomorphic design cues and communicative agency framing on conversational agent and company perceptions}.
\newblock \bibinfo{journal}{\emph{Computers in Human Behavior}}  \bibinfo{volume}{85} (\bibinfo{year}{2018}), \bibinfo{pages}{183--189}.
\newblock
\showISSN{0747-5632}
\href{https://doi.org/10.1016/j.chb.2018.03.051}{doi:\nolinkurl{10.1016/j.chb.2018.03.051}}


\bibitem[Baraka and Veloso(2018)]%
        {baraka2018mobile}
\bibfield{author}{\bibinfo{person}{Kim Baraka} {and} \bibinfo{person}{Manuela~M Veloso}.} \bibinfo{year}{2018}\natexlab{}.
\newblock \showarticletitle{Mobile service robot state revealing through expressive lights: formalism, design, and evaluation}.
\newblock \bibinfo{journal}{\emph{International Journal of Social Robotics}}  \bibinfo{volume}{10} (\bibinfo{year}{2018}), \bibinfo{pages}{65--92}.
\newblock


\bibitem[Bartholomew and Mehta(2023)]%
        {bartholomew2023}
\bibfield{author}{\bibinfo{person}{Jem Bartholomew} {and} \bibinfo{person}{Dhrumil Mehta}.} \bibinfo{year}{2023}\natexlab{}.
\newblock \bibinfo{title}{The Business of Artificial Intelligence}.
\newblock \bibinfo{howpublished}{Columbia Journalism Review}.
\newblock
\urldef\tempurl%
\url{https://www.cjr.org/tow_center/media-coverage-chatgpt.php}
\showURL{%
\tempurl}


\bibitem[Bartholomew and Mehta(2024)]%
        {bloomberg2024}
\bibfield{author}{\bibinfo{person}{Jem Bartholomew} {and} \bibinfo{person}{Dhrumil Mehta}.} \bibinfo{year}{2024}\natexlab{}.
\newblock \bibinfo{title}{Big tech 2025 capex may hit \$200 billion as gen-AI demand booms}.
\newblock \bibinfo{howpublished}{Bloomberg Intelligence}.
\newblock
\urldef\tempurl%
\url{https://www.bloomberg.com/professional/insights/technology/big-tech-2025-capex-may-hit-200-billion-as-gen-ai-demand-booms/}
\showURL{%
\tempurl}


\bibitem[Bennett et~al\mbox{.}(2024)]%
        {bennett2024}
\bibfield{author}{\bibinfo{person}{Cynthia~L. Bennett}, \bibinfo{person}{Renee Shelby}, \bibinfo{person}{Negar Rostamzadeh}, {and} \bibinfo{person}{Shaun~K Kane}.} \bibinfo{year}{2024}\natexlab{}.
\newblock \showarticletitle{Painting with Cameras and Drawing with Text: AI Use in Accessible Creativity}. In \bibinfo{booktitle}{\emph{Proceedings of the 26th International ACM SIGACCESS Conference on Computers and Accessibility}} (St. John's, NL, Canada) \emph{(\bibinfo{series}{ASSETS '24})}. \bibinfo{publisher}{Association for Computing Machinery}, \bibinfo{address}{New York, NY, USA}, Article \bibinfo{articleno}{5}, \bibinfo{numpages}{19}~pages.
\newblock
\showISBNx{9798400706776}
\href{https://doi.org/10.1145/3663548.3675644}{doi:\nolinkurl{10.1145/3663548.3675644}}


\bibitem[Berney et~al\mbox{.}(2024)]%
        {berney2024care}
\bibfield{author}{\bibinfo{person}{Manon Berney}, \bibinfo{person}{Abdessalam Ouaazki}, \bibinfo{person}{Vladimir Macko}, \bibinfo{person}{Bruno Kocher}, {and} \bibinfo{person}{Adrian Holzer}.} \bibinfo{year}{2024}\natexlab{}.
\newblock \showarticletitle{Care-Based Eco-Feedback Augmented with Generative AI: Fostering Pro-Environmental Behavior through Emotional Attachment}. In \bibinfo{booktitle}{\emph{Proceedings of the 2024 CHI Conference on Human Factors in Computing Systems}} (Honolulu, HI, USA) \emph{(\bibinfo{series}{CHI '24})}. \bibinfo{publisher}{Association for Computing Machinery}, \bibinfo{address}{New York, NY, USA}, Article \bibinfo{articleno}{469}, \bibinfo{numpages}{15}~pages.
\newblock
\showISBNx{9798400703300}
\href{https://doi.org/10.1145/3613904.3642296}{doi:\nolinkurl{10.1145/3613904.3642296}}


\bibitem[Berry et~al\mbox{.}(2005)]%
        {berry2005evaluating}
\bibfield{author}{\bibinfo{person}{Dianne~C Berry}, \bibinfo{person}{Laurie~T Butler}, {and} \bibinfo{person}{Fiorella De~Rosis}.} \bibinfo{year}{2005}\natexlab{}.
\newblock \showarticletitle{Evaluating a realistic agent in an advice-giving task}.
\newblock \bibinfo{journal}{\emph{International Journal of Human-Computer Studies}} \bibinfo{volume}{63}, \bibinfo{number}{3} (\bibinfo{year}{2005}), \bibinfo{pages}{304--327}.
\newblock


\bibitem[Bi and Huang(2023)]%
        {bi2023}
\bibfield{author}{\bibinfo{person}{Nanyi Bi} {and} \bibinfo{person}{Janet Yi-Ching Huang}.} \bibinfo{year}{2023}\natexlab{}.
\newblock \showarticletitle{I create, therefore I agree: Exploring the effect of AI anthropomorphism on human decision-making}. In \bibinfo{booktitle}{\emph{Companion Publication of the 2023 Conference on Computer Supported Cooperative Work and Social Computing}} (Minneapolis, MN, USA) \emph{(\bibinfo{series}{CSCW '23 Companion})}. \bibinfo{publisher}{Association for Computing Machinery}, \bibinfo{address}{New York, NY, USA}, \bibinfo{pages}{241–244}.
\newblock
\showISBNx{9798400701290}
\href{https://doi.org/10.1145/3584931.3606990}{doi:\nolinkurl{10.1145/3584931.3606990}}


\bibitem[Bickmore and Cassell(2005)]%
        {bickmore2005social}
\bibfield{author}{\bibinfo{person}{Timothy Bickmore} {and} \bibinfo{person}{Justine Cassell}.} \bibinfo{year}{2005}\natexlab{}.
\newblock \bibinfo{booktitle}{\emph{Social Dialongue with Embodied Conversational Agents}}.
\newblock \bibinfo{publisher}{Springer Netherlands}, \bibinfo{address}{Dordrecht}, \bibinfo{pages}{23--54}.
\newblock
\showISBNx{978-1-4020-3933-1}
\href{https://doi.org/10.1007/1-4020-3933-6_2}{doi:\nolinkurl{10.1007/1-4020-3933-6_2}}


\bibitem[Boyarskaya et~al\mbox{.}(2020)]%
        {boyarskaya2020}
\bibfield{author}{\bibinfo{person}{Margarita Boyarskaya}, \bibinfo{person}{Alexandra Olteanu}, {and} \bibinfo{person}{Kate Crawford}.} \bibinfo{year}{2020}\natexlab{}.
\newblock \bibinfo{title}{Overcoming Failures of Imagination in AI Infused System Development and Deployment}.
\newblock
\showeprint[arxiv]{2011.13416}~[cs.CY]
\urldef\tempurl%
\url{https://arxiv.org/abs/2011.13416}
\showURL{%
\tempurl}


\bibitem[Brandtzaeg et~al\mbox{.}(2022)]%
        {brandtzaeg2022}
\bibfield{author}{\bibinfo{person}{Petter~Bae Brandtzaeg}, \bibinfo{person}{Marita Skjuve}, {and} \bibinfo{person}{Asbjørn Følstad}.} \bibinfo{year}{2022}\natexlab{}.
\newblock \showarticletitle{{My AI Friend: How Users of a Social Chatbot Understand Their Human–AI Friendship}}.
\newblock \bibinfo{journal}{\emph{Human Communication Research}} \bibinfo{volume}{48}, \bibinfo{number}{3} (\bibinfo{date}{04} \bibinfo{year}{2022}), \bibinfo{pages}{404--429}.
\newblock
\showISSN{1468-2958}
\showeprint{https://academic.oup.com/hcr/article-pdf/48/3/404/44308072/hqac008.pdf}
\href{https://doi.org/10.1093/hcr/hqac008}{doi:\nolinkurl{10.1093/hcr/hqac008}}


\bibitem[Briggs and Kodnani(2023)]%
        {briggs2023}
\bibfield{author}{\bibinfo{person}{Joseph Briggs} {and} \bibinfo{person}{Devesh Kodnani}.} \bibinfo{year}{2023}\natexlab{}.
\newblock \bibinfo{booktitle}{\emph{The Potentially Large Effects of Artificial Intelligence on Economic Growth}}.
\newblock \bibinfo{type}{Global Economics Analyst}. \bibinfo{institution}{Goldman Sachs}.
\newblock


\bibitem[Bubeck et~al\mbox{.}(2023)]%
        {bubeck2023sparks}
\bibfield{author}{\bibinfo{person}{S{\'e}bastien Bubeck}, \bibinfo{person}{Varun Chandrasekaran}, \bibinfo{person}{Ronen Eldan}, \bibinfo{person}{Johannes Gehrke}, \bibinfo{person}{Eric Horvitz}, \bibinfo{person}{Ece Kamar}, \bibinfo{person}{Peter Lee}, \bibinfo{person}{Yin~Tat Lee}, \bibinfo{person}{Yuanzhi Li}, \bibinfo{person}{Scott Lundberg}, {et~al\mbox{.}}} \bibinfo{year}{2023}\natexlab{}.
\newblock \showarticletitle{Sparks of artificial general intelligence: Early experiments with gpt-4}.
\newblock \bibinfo{journal}{\emph{arXiv preprint arXiv:2303.12712}} (\bibinfo{year}{2023}).
\newblock


\bibitem[Burrell(2016)]%
        {burrell2016}
\bibfield{author}{\bibinfo{person}{Jenna Burrell}.} \bibinfo{year}{2016}\natexlab{}.
\newblock \showarticletitle{How the machine ‘thinks’: Understanding opacity in machine learning algorithms}.
\newblock \bibinfo{journal}{\emph{Big Data \& Society}} \bibinfo{volume}{3}, \bibinfo{number}{1} (\bibinfo{year}{2016}), \bibinfo{pages}{2053951715622512}.
\newblock
\showeprint{https://doi.org/10.1177/2053951715622512}
\href{https://doi.org/10.1177/2053951715622512}{doi:\nolinkurl{10.1177/2053951715622512}}


\bibitem[Carlsen and Glenton(2011)]%
        {carlsen2011n}
\bibfield{author}{\bibinfo{person}{Benedicte Carlsen} {and} \bibinfo{person}{Claire Glenton}.} \bibinfo{year}{2011}\natexlab{}.
\newblock \showarticletitle{What about N? A methodological study of sample-size reporting in focus group studies}.
\newblock \bibinfo{journal}{\emph{BMC medical research methodology}} \bibinfo{volume}{11}, \bibinfo{number}{1} (\bibinfo{year}{2011}), \bibinfo{pages}{26}.
\newblock


\bibitem[Catrambone et~al\mbox{.}(2019)]%
        {catrambone2019anthropomorphic}
\bibfield{author}{\bibinfo{person}{Richard Catrambone}, \bibinfo{person}{John Stasko}, {and} \bibinfo{person}{Jun Xiao}.} \bibinfo{year}{2019}\natexlab{}.
\newblock \showarticletitle{Anthropomorphic agents as a user interface paradigm: Experimental findings and a framework for research}. In \bibinfo{booktitle}{\emph{Proceedings of the Twenty-Fourth Annual Conference of the Cognitive Science Society}}. Routledge, \bibinfo{publisher}{Cognitive Science Society}, \bibinfo{address}{Fairfax, Virginia, USA}, \bibinfo{pages}{166--171}.
\newblock


\bibitem[Chancellor(2023)]%
        {chancellor2023toward}
\bibfield{author}{\bibinfo{person}{Stevie Chancellor}.} \bibinfo{year}{2023}\natexlab{}.
\newblock \showarticletitle{Toward practices for human-centered machine learning}.
\newblock \bibinfo{journal}{\emph{Commun. ACM}} \bibinfo{volume}{66}, \bibinfo{number}{3} (\bibinfo{year}{2023}), \bibinfo{pages}{78--85}.
\newblock


\bibitem[Chandra et~al\mbox{.}(2022)]%
        {chandra2022or}
\bibfield{author}{\bibinfo{person}{Shalini Chandra}, \bibinfo{person}{Anuragini Shirish}, {and} \bibinfo{person}{Shirish~C Srivastava}.} \bibinfo{year}{2022}\natexlab{}.
\newblock \showarticletitle{To be or not to be… human? Theorizing the role of human-like competencies in conversational artificial intelligence agents}.
\newblock \bibinfo{journal}{\emph{Journal of Management Information Systems}} \bibinfo{volume}{39}, \bibinfo{number}{4} (\bibinfo{year}{2022}), \bibinfo{pages}{969--1005}.
\newblock


\bibitem[Character.AI(2024)]%
        {character2024}
\bibfield{author}{\bibinfo{person}{Character.AI}.} \bibinfo{year}{2024}\natexlab{}.
\newblock \bibinfo{title}{Character.AI}.
\newblock
\urldef\tempurl%
\url{https://character.ai//}
\showURL{%
\tempurl}


\bibitem[Chattaraman et~al\mbox{.}(2019)]%
        {chattaraman2019should}
\bibfield{author}{\bibinfo{person}{Veena Chattaraman}, \bibinfo{person}{Wi-Suk Kwon}, \bibinfo{person}{Juan~E Gilbert}, {and} \bibinfo{person}{Kassandra Ross}.} \bibinfo{year}{2019}\natexlab{}.
\newblock \showarticletitle{Should AI-Based, conversational digital assistants employ social-or task-oriented interaction style? A task-competency and reciprocity perspective for older adults}.
\newblock \bibinfo{journal}{\emph{Computers in Human Behavior}}  \bibinfo{volume}{90} (\bibinfo{year}{2019}), \bibinfo{pages}{315--330}.
\newblock


\bibitem[Cheng et~al\mbox{.}(2024)]%
        {cheng2024one}
\bibfield{author}{\bibinfo{person}{Myra Cheng}, \bibinfo{person}{Alicia DeVrio}, \bibinfo{person}{Lisa Egede}, \bibinfo{person}{Su~Lin Blodgett}, {and} \bibinfo{person}{Alexandra Olteanu}.} \bibinfo{year}{2024}\natexlab{}.
\newblock \bibinfo{title}{"I Am the One and Only, Your Cyber BFF": Understanding the Impact of GenAI Requires Understanding the Impact of Anthropomorphic AI}.
\newblock


\bibitem[Chomsky(2023)]%
        {chomsky2023}
\bibfield{author}{\bibinfo{person}{Noam Chomsky}.} \bibinfo{year}{2023}\natexlab{}.
\newblock \bibinfo{title}{Opinion Guest Essay: The False Promise of ChatGPT}.
\newblock
\urldef\tempurl%
\url{https://www.nytimes.com/2023/03/08/opinion/noam-chomsky-chatgpt-ai.html}
\showURL{%
\tempurl}


\bibitem[Christoforakos and Diefenbach(2023)]%
        {christoforakos2023technology}
\bibfield{author}{\bibinfo{person}{Lara Christoforakos} {and} \bibinfo{person}{Sarah Diefenbach}.} \bibinfo{year}{2023}\natexlab{}.
\newblock \showarticletitle{Technology as a social companion? An exploration of individual and product-related factors of anthropomorphism}.
\newblock \bibinfo{journal}{\emph{Social Science Computer Review}} \bibinfo{volume}{41}, \bibinfo{number}{3} (\bibinfo{year}{2023}), \bibinfo{pages}{1039--1062}.
\newblock


\bibitem[Cooper et~al\mbox{.}(2022)]%
        {cooper2022}
\bibfield{author}{\bibinfo{person}{A.~Feder Cooper}, \bibinfo{person}{Emanuel Moss}, \bibinfo{person}{Benjamin Laufer}, {and} \bibinfo{person}{Helen Nissenbaum}.} \bibinfo{year}{2022}\natexlab{}.
\newblock \showarticletitle{Accountability in an Algorithmic Society: Relationality, Responsibility, and Robustness in Machine Learning}. In \bibinfo{booktitle}{\emph{Proceedings of the 2022 ACM Conference on Fairness, Accountability, and Transparency}} (Seoul, Republic of Korea) \emph{(\bibinfo{series}{FAccT '22})}. \bibinfo{publisher}{Association for Computing Machinery}, \bibinfo{address}{New York, NY, USA}, \bibinfo{pages}{864–876}.
\newblock
\showISBNx{9781450393522}
\href{https://doi.org/10.1145/3531146.3533150}{doi:\nolinkurl{10.1145/3531146.3533150}}


\bibitem[{Cooper}(2000)]%
        {cooper2000}
\bibfield{author}{\bibinfo{person}{M.D. {Cooper}}.} \bibinfo{year}{2000}\natexlab{}.
\newblock \showarticletitle{Towards a model of safety culture}.
\newblock \bibinfo{journal}{\emph{Safety Science}} \bibinfo{volume}{36}, \bibinfo{number}{2} (\bibinfo{year}{2000}), \bibinfo{pages}{111--136}.
\newblock
\showISSN{0925-7535}
\href{https://doi.org/10.1016/S0925-7535(00)00035-7}{doi:\nolinkurl{10.1016/S0925-7535(00)00035-7}}


\bibitem[Cowell and Stanney(2005)]%
        {cowell2005manipulation}
\bibfield{author}{\bibinfo{person}{Andrew~J Cowell} {and} \bibinfo{person}{Kay~M Stanney}.} \bibinfo{year}{2005}\natexlab{}.
\newblock \showarticletitle{Manipulation of non-verbal interaction style and demographic embodiment to increase anthropomorphic computer character credibility}.
\newblock \bibinfo{journal}{\emph{International journal of human-computer studies}} \bibinfo{volume}{62}, \bibinfo{number}{2} (\bibinfo{year}{2005}), \bibinfo{pages}{281--306}.
\newblock


\bibitem[Cox et~al\mbox{.}(2016)]%
        {cox2016}
\bibfield{author}{\bibinfo{person}{Anna~L. Cox}, \bibinfo{person}{Sandy~J.J. Gould}, \bibinfo{person}{Marta~E. Cecchinato}, \bibinfo{person}{Ioanna Iacovides}, {and} \bibinfo{person}{Ian Renfree}.} \bibinfo{year}{2016}\natexlab{}.
\newblock \showarticletitle{Design Frictions for Mindful Interactions: The Case for Microboundaries}. In \bibinfo{booktitle}{\emph{Proceedings of the 2016 CHI Conference Extended Abstracts on Human Factors in Computing Systems}} (San Jose, California, USA) \emph{(\bibinfo{series}{CHI EA '16})}. \bibinfo{publisher}{Association for Computing Machinery}, \bibinfo{address}{New York, NY, USA}, \bibinfo{pages}{1389–1397}.
\newblock
\showISBNx{9781450340823}
\href{https://doi.org/10.1145/2851581.2892410}{doi:\nolinkurl{10.1145/2851581.2892410}}


\bibitem[Culley and Madhavan(2013)]%
        {CULLEY2013577}
\bibfield{author}{\bibinfo{person}{Kimberly~E. Culley} {and} \bibinfo{person}{Poornima Madhavan}.} \bibinfo{year}{2013}\natexlab{}.
\newblock \showarticletitle{A note of caution regarding anthropomorphism in HCI agents}.
\newblock \bibinfo{journal}{\emph{Computers in Human Behavior}} \bibinfo{volume}{29}, \bibinfo{number}{3} (\bibinfo{year}{2013}), \bibinfo{pages}{577--579}.
\newblock
\showISSN{0747-5632}
\href{https://doi.org/10.1016/j.chb.2012.11.023}{doi:\nolinkurl{10.1016/j.chb.2012.11.023}}


\bibitem[Davis(2023)]%
        {davis2023}
\bibfield{author}{\bibinfo{person}{Jenny~L. Davis}.} \bibinfo{year}{2023}\natexlab{}.
\newblock \showarticletitle{‘Affordances’ for Machine Learning}. In \bibinfo{booktitle}{\emph{Proceedings of the 2023 ACM Conference on Fairness, Accountability, and Transparency}} (Chicago, IL, USA) \emph{(\bibinfo{series}{FAccT '23})}. \bibinfo{publisher}{Association for Computing Machinery}, \bibinfo{address}{New York, NY, USA}, \bibinfo{pages}{324–332}.
\newblock
\showISBNx{9798400701924}
\href{https://doi.org/10.1145/3593013.3594000}{doi:\nolinkurl{10.1145/3593013.3594000}}


\bibitem[De~Visser et~al\mbox{.}(2016)]%
        {de2016almost}
\bibfield{author}{\bibinfo{person}{Ewart~J. De~Visser}, \bibinfo{person}{Samuel~S. Monfort}, \bibinfo{person}{Ryan McKendrick}, \bibinfo{person}{Melissa~A.B. Smith}, \bibinfo{person}{Patrick~E. McKnight}, \bibinfo{person}{Frank Krueger}, {and} \bibinfo{person}{Raja Parasuraman}.} \bibinfo{year}{2016}\natexlab{}.
\newblock \showarticletitle{Almost human: Anthropomorphism increases trust resilience in cognitive agents.}
\newblock \bibinfo{journal}{\emph{Journal of Experimental Psychology: Applied}} \bibinfo{volume}{22}, \bibinfo{number}{3} (\bibinfo{year}{2016}), \bibinfo{pages}{331}.
\newblock


\bibitem[Derks et~al\mbox{.}(2008)]%
        {derks2008emoticons}
\bibfield{author}{\bibinfo{person}{Daantje Derks}, \bibinfo{person}{Arjan~ER Bos}, {and} \bibinfo{person}{Jasper Von~Grumbkow}.} \bibinfo{year}{2008}\natexlab{}.
\newblock \showarticletitle{Emoticons in computer-mediated communication: Social motives and social context}.
\newblock \bibinfo{journal}{\emph{Cyberpsychology \& behavior}} \bibinfo{volume}{11}, \bibinfo{number}{1} (\bibinfo{year}{2008}), \bibinfo{pages}{99--101}.
\newblock


\bibitem[Dobbe and Wolters(2024)]%
        {dobbe2024toward}
\bibfield{author}{\bibinfo{person}{Roel Dobbe} {and} \bibinfo{person}{Anouk Wolters}.} \bibinfo{year}{2024}\natexlab{}.
\newblock \showarticletitle{Toward Sociotechnical AI: Mapping Vulnerabilities for Machine Learning in Context}.
\newblock \bibinfo{journal}{\emph{Minds and Machines}} \bibinfo{volume}{34}, \bibinfo{number}{2} (\bibinfo{year}{2024}), \bibinfo{pages}{1--51}.
\newblock


\bibitem[Don et~al\mbox{.}(1992)]%
        {don1992anthropomorphism}
\bibfield{author}{\bibinfo{person}{Abbe Don}, \bibinfo{person}{Susan Brennan}, \bibinfo{person}{Brenda Laurel}, {and} \bibinfo{person}{Ben Shneiderman}.} \bibinfo{year}{1992}\natexlab{}.
\newblock \showarticletitle{Anthropomorphism: from Eliza to Terminator 2}. In \bibinfo{booktitle}{\emph{Proceedings of the SIGCHI Conference on Human Factors in Computing Systems}} (Monterey, California, USA) \emph{(\bibinfo{series}{CHI '92})}. \bibinfo{publisher}{Association for Computing Machinery}, \bibinfo{address}{New York, NY, USA}, \bibinfo{pages}{67–70}.
\newblock
\showISBNx{0897915135}
\href{https://doi.org/10.1145/142750.142760}{doi:\nolinkurl{10.1145/142750.142760}}


\bibitem[Eloundou et~al\mbox{.}(2023)]%
        {eloundou2023}
\bibfield{author}{\bibinfo{person}{Tyna Eloundou}, \bibinfo{person}{Sam Manning}, \bibinfo{person}{Pamela Mishkin}, {and} \bibinfo{person}{Daniel Rock}.} \bibinfo{year}{2023}\natexlab{}.
\newblock \bibinfo{title}{GPTs are GPTs: An Early Look at the Labor Market Impact Potential of Large Language Models}.
\newblock
\showeprint[arxiv]{2303.10130}~[econ.GN]


\bibitem[Epley et~al\mbox{.}(2008)]%
        {epley2008we}
\bibfield{author}{\bibinfo{person}{Nicholas Epley}, \bibinfo{person}{Adam Waytz}, \bibinfo{person}{Scott Akalis}, {and} \bibinfo{person}{John~T Cacioppo}.} \bibinfo{year}{2008}\natexlab{}.
\newblock \showarticletitle{When we need a human: Motivational determinants of anthropomorphism}.
\newblock \bibinfo{journal}{\emph{Social cognition}} \bibinfo{volume}{26}, \bibinfo{number}{2} (\bibinfo{year}{2008}), \bibinfo{pages}{143--155}.
\newblock


\bibitem[Epley et~al\mbox{.}(2007)]%
        {epley2007seeing}
\bibfield{author}{\bibinfo{person}{Nicholas Epley}, \bibinfo{person}{Adam Waytz}, {and} \bibinfo{person}{John~T Cacioppo}.} \bibinfo{year}{2007}\natexlab{}.
\newblock \showarticletitle{On seeing human: a three-factor theory of anthropomorphism.}
\newblock \bibinfo{journal}{\emph{Psychological review}} \bibinfo{volume}{114}, \bibinfo{number}{4} (\bibinfo{year}{2007}), \bibinfo{pages}{864}.
\newblock


\bibitem[Feine et~al\mbox{.}(2019)]%
        {feine2019taxonomy}
\bibfield{author}{\bibinfo{person}{Jasper Feine}, \bibinfo{person}{Ulrich Gnewuch}, \bibinfo{person}{Stefan Morana}, {and} \bibinfo{person}{Alexander Maedche}.} \bibinfo{year}{2019}\natexlab{}.
\newblock \showarticletitle{A taxonomy of social cues for conversational agents}.
\newblock \bibinfo{journal}{\emph{International Journal of human-computer studies}}  \bibinfo{volume}{132} (\bibinfo{year}{2019}), \bibinfo{pages}{138--161}.
\newblock


\bibitem[Fink(2012)]%
        {fink2012anthropomorphism}
\bibfield{author}{\bibinfo{person}{Julia Fink}.} \bibinfo{year}{2012}\natexlab{}.
\newblock \showarticletitle{Anthropomorphism and Human Likeness in the Design of Robots and Human-Robot Interaction}. In \bibinfo{booktitle}{\emph{Social Robotics}}, \bibfield{editor}{\bibinfo{person}{Shuzhi~Sam Ge}, \bibinfo{person}{Oussama Khatib}, \bibinfo{person}{John-John Cabibihan}, \bibinfo{person}{Reid Simmons}, {and} \bibinfo{person}{Mary-Anne Williams}} (Eds.). \bibinfo{publisher}{Springer Berlin Heidelberg}, \bibinfo{address}{Berlin, Heidelberg}, \bibinfo{pages}{199--208}.
\newblock
\showISBNx{978-3-642-34103-8}


\bibitem[Fox et~al\mbox{.}(2023)]%
        {fox2023}
\bibfield{author}{\bibinfo{person}{Sarah~E. Fox}, \bibinfo{person}{Samantha Shorey}, \bibinfo{person}{Esther~Y. Kang}, \bibinfo{person}{Dominique Montiel~Valle}, {and} \bibinfo{person}{Estefania Rodriguez}.} \bibinfo{year}{2023}\natexlab{}.
\newblock \showarticletitle{Patchwork: The Hidden, Human Labor of AI Integration within Essential Work}.
\newblock \bibinfo{journal}{\emph{Proc. ACM Hum.-Comput. Interact.}} \bibinfo{volume}{7}, \bibinfo{number}{CSCW1}, Article \bibinfo{articleno}{81} (\bibinfo{date}{April} \bibinfo{year}{2023}), \bibinfo{numpages}{20}~pages.
\newblock
\href{https://doi.org/10.1145/3579514}{doi:\nolinkurl{10.1145/3579514}}


\bibitem[Gabriel et~al\mbox{.}(2024)]%
        {gabriel2024ethics}
\bibfield{author}{\bibinfo{person}{Iason Gabriel}, \bibinfo{person}{Arianna Manzini}, \bibinfo{person}{Geoff Keeling}, \bibinfo{person}{Lisa~Anne Hendricks}, \bibinfo{person}{Verena Rieser}, \bibinfo{person}{Hasan Iqbal}, \bibinfo{person}{Nenad Toma{\v{s}}ev}, \bibinfo{person}{Ira Ktena}, \bibinfo{person}{Zachary Kenton}, \bibinfo{person}{Mikel Rodriguez}, {et~al\mbox{.}}} \bibinfo{year}{2024}\natexlab{}.
\newblock \bibinfo{title}{The Ethics of Advanced AI Assistants}.
\newblock


\bibitem[Gambino et~al\mbox{.}(2020)]%
        {gambino2020building}
\bibfield{author}{\bibinfo{person}{Andrew Gambino}, \bibinfo{person}{Jesse Fox}, {and} \bibinfo{person}{Rabindra~A. Ratan}.} \bibinfo{year}{2020}\natexlab{}.
\newblock \showarticletitle{Building a stronger CASA: Extending the computers are social actors paradigm}.
\newblock \bibinfo{journal}{\emph{Human-Machine Communication}}  \bibinfo{volume}{1} (\bibinfo{year}{2020}), \bibinfo{pages}{71--85}.
\newblock


\bibitem[Geerdts(2016)]%
        {geerdts2016real}
\bibfield{author}{\bibinfo{person}{Megan~S. Geerdts}.} \bibinfo{year}{2016}\natexlab{}.
\newblock \showarticletitle{(Un) real animals: Anthropomorphism and early learning about animals}.
\newblock \bibinfo{journal}{\emph{Child Development Perspectives}} \bibinfo{volume}{10}, \bibinfo{number}{1} (\bibinfo{year}{2016}), \bibinfo{pages}{10--14}.
\newblock


\bibitem[Glikson and Woolley(2020)]%
        {glikson2020human}
\bibfield{author}{\bibinfo{person}{Ella Glikson} {and} \bibinfo{person}{Anita~Williams Woolley}.} \bibinfo{year}{2020}\natexlab{}.
\newblock \showarticletitle{Human trust in artificial intelligence: Review of empirical research}.
\newblock \bibinfo{journal}{\emph{Academy of management annals}} \bibinfo{volume}{14}, \bibinfo{number}{2} (\bibinfo{year}{2020}), \bibinfo{pages}{627--660}.
\newblock


\bibitem[Gnewuch et~al\mbox{.}(2018)]%
        {gnewuch2018faster}
\bibfield{author}{\bibinfo{person}{Ulrich Gnewuch}, \bibinfo{person}{Stefan Morana}, \bibinfo{person}{Marc~TP Adam}, {and} \bibinfo{person}{Alexander Maedche}.} \bibinfo{year}{2018}\natexlab{}.
\newblock \showarticletitle{Faster Is Not Always Better: Understanding the Effect of Dynamic Response Delays in Human-Chatbot Interaction}. In \bibinfo{booktitle}{\emph{26th European Conference on Information Systems: Beyond Digitization-Facets of Socio-Technical Change, ECIS 2018, Portsmouth, UK, June 23-28, 2018. Ed.: U. Frank}}. \bibinfo{publisher}{ECIS}, \bibinfo{address}{Portsmouth, UK}, \bibinfo{pages}{143975}.
\newblock


\bibitem[Gong(2008)]%
        {gong2008social}
\bibfield{author}{\bibinfo{person}{Li Gong}.} \bibinfo{year}{2008}\natexlab{}.
\newblock \showarticletitle{How social is social responses to computers? The function of the degree of anthropomorphism in computer representations}.
\newblock \bibinfo{journal}{\emph{Computers in Human Behavior}} \bibinfo{volume}{24}, \bibinfo{number}{4} (\bibinfo{year}{2008}), \bibinfo{pages}{1494--1509}.
\newblock


\bibitem[Green and Viljoen(2020)]%
        {green2020}
\bibfield{author}{\bibinfo{person}{Ben Green} {and} \bibinfo{person}{Salom\'{e} Viljoen}.} \bibinfo{year}{2020}\natexlab{}.
\newblock \showarticletitle{Algorithmic realism: expanding the boundaries of algorithmic thought}. In \bibinfo{booktitle}{\emph{Proceedings of the 2020 Conference on Fairness, Accountability, and Transparency}} (Barcelona, Spain) \emph{(\bibinfo{series}{FAT* '20})}. \bibinfo{publisher}{Association for Computing Machinery}, \bibinfo{address}{New York, NY, USA}, \bibinfo{pages}{19–31}.
\newblock
\showISBNx{9781450369367}
\href{https://doi.org/10.1145/3351095.3372840}{doi:\nolinkurl{10.1145/3351095.3372840}}


\bibitem[Guest et~al\mbox{.}(2017)]%
        {guest2017many}
\bibfield{author}{\bibinfo{person}{Greg Guest}, \bibinfo{person}{Emily Namey}, {and} \bibinfo{person}{Kevin McKenna}.} \bibinfo{year}{2017}\natexlab{}.
\newblock \showarticletitle{How many focus groups are enough? Building an evidence base for nonprobability sample sizes}.
\newblock \bibinfo{journal}{\emph{Field methods}} \bibinfo{volume}{29}, \bibinfo{number}{1} (\bibinfo{year}{2017}), \bibinfo{pages}{3--22}.
\newblock


\bibitem[Guthrie(1995)]%
        {guthrie1995faces}
\bibfield{author}{\bibinfo{person}{Stewart~Elliott Guthrie}.} \bibinfo{year}{1995}\natexlab{}.
\newblock \bibinfo{booktitle}{\emph{Faces in the Clouds: A New Theory of Religion}}.
\newblock \bibinfo{publisher}{Oxford University Press}, \bibinfo{address}{New York, New York, USA}.
\newblock


\bibitem[Ha et~al\mbox{.}(2021)]%
        {ha2021exploring}
\bibfield{author}{\bibinfo{person}{Quang-An Ha}, \bibinfo{person}{Jengchung~Victor Chen}, \bibinfo{person}{Ha~Uy Uy}, {and} \bibinfo{person}{Erik~Paolo Capistrano}.} \bibinfo{year}{2021}\natexlab{}.
\newblock \showarticletitle{Exploring the privacy concerns in using intelligent virtual assistants under perspectives of information sensitivity and anthropomorphism}.
\newblock \bibinfo{journal}{\emph{International journal of human--computer interaction}} \bibinfo{volume}{37}, \bibinfo{number}{6} (\bibinfo{year}{2021}), \bibinfo{pages}{512--527}.
\newblock


\bibitem[Hales(1994)]%
        {Hales1994}
\bibfield{author}{\bibinfo{person}{M. Hales}.} \bibinfo{year}{1994}\natexlab{}.
\newblock \bibinfo{booktitle}{\emph{Design Issues in CSCW}}.
\newblock \bibinfo{publisher}{Springer London}, \bibinfo{address}{London}, Chapter Where Are Designers? Styles of Design Practice, Objects of Design and Views of Users in CSCW, \bibinfo{pages}{151--177}.
\newblock
\showISBNx{978-1-4471-2029-2}
\href{https://doi.org/10.1007/978-1-4471-2029-2_8}{doi:\nolinkurl{10.1007/978-1-4471-2029-2_8}}


\bibitem[Heger et~al\mbox{.}(2022)]%
        {heger2022understanding}
\bibfield{author}{\bibinfo{person}{Amy~K Heger}, \bibinfo{person}{Liz~B Marquis}, \bibinfo{person}{Mihaela Vorvoreanu}, \bibinfo{person}{Hanna Wallach}, {and} \bibinfo{person}{Jennifer Wortman~Vaughan}.} \bibinfo{year}{2022}\natexlab{}.
\newblock \showarticletitle{Understanding machine learning practitioners' data documentation perceptions, needs, challenges, and desiderata}.
\newblock \bibinfo{journal}{\emph{Proceedings of the ACM on Human-Computer Interaction}} \bibinfo{volume}{6}, \bibinfo{number}{CSCW2} (\bibinfo{year}{2022}), \bibinfo{pages}{1--29}.
\newblock


\bibitem[Holstein et~al\mbox{.}(2020)]%
        {holstein2020conceptual}
\bibfield{author}{\bibinfo{person}{Kenneth Holstein}, \bibinfo{person}{Vincent Aleven}, {and} \bibinfo{person}{Nikol Rummel}.} \bibinfo{year}{2020}\natexlab{}.
\newblock \showarticletitle{A Conceptual Framework for Human--AI Hybrid Adaptivity in Education}. In \bibinfo{booktitle}{\emph{Artificial Intelligence in Education}}, \bibfield{editor}{\bibinfo{person}{Ig~Ibert Bittencourt}, \bibinfo{person}{Mutlu Cukurova}, \bibinfo{person}{Kasia Muldner}, \bibinfo{person}{Rose Luckin}, {and} \bibinfo{person}{Eva Mill{\'a}n}} (Eds.). \bibinfo{publisher}{Springer International Publishing}, \bibinfo{address}{Cham}, \bibinfo{pages}{240--254}.
\newblock
\showISBNx{978-3-030-52237-7}


\bibitem[Holstein et~al\mbox{.}(2019)]%
        {holstein2019improving}
\bibfield{author}{\bibinfo{person}{Kenneth Holstein}, \bibinfo{person}{Jennifer Wortman~Vaughan}, \bibinfo{person}{Hal Daum\'{e}}, \bibinfo{person}{Miro Dudik}, {and} \bibinfo{person}{Hanna Wallach}.} \bibinfo{year}{2019}\natexlab{}.
\newblock \showarticletitle{Improving Fairness in Machine Learning Systems: What Do Industry Practitioners Need?}. In \bibinfo{booktitle}{\emph{Proceedings of the 2019 CHI Conference on Human Factors in Computing Systems}} (Glasgow, Scotland Uk) \emph{(\bibinfo{series}{CHI '19})}. \bibinfo{publisher}{Association for Computing Machinery}, \bibinfo{address}{New York, NY, USA}, \bibinfo{pages}{1–16}.
\newblock
\showISBNx{9781450359702}
\href{https://doi.org/10.1145/3290605.3300830}{doi:\nolinkurl{10.1145/3290605.3300830}}


\bibitem[Hurshman(2024)]%
        {hurshman2024opaque}
\bibfield{author}{\bibinfo{person}{Clint Hurshman}.} \bibinfo{year}{2024}\natexlab{}.
\newblock \showarticletitle{Do opaque algorithms have functions?}
\newblock \bibinfo{journal}{\emph{Synthese}} \bibinfo{volume}{204}, \bibinfo{number}{3} (\bibinfo{year}{2024}), \bibinfo{pages}{91}.
\newblock


\bibitem[Inie et~al\mbox{.}(2024)]%
        {inie2024}
\bibfield{author}{\bibinfo{person}{Nanna Inie}, \bibinfo{person}{Stefania Druga}, \bibinfo{person}{Peter Zukerman}, {and} \bibinfo{person}{Emily~M. Bender}.} \bibinfo{year}{2024}\natexlab{}.
\newblock \showarticletitle{From "AI" to Probabilistic Automation: How Does Anthropomorphization of Technical Systems Descriptions Influence Trust?}. In \bibinfo{booktitle}{\emph{Proceedings of the 2024 ACM Conference on Fairness, Accountability, and Transparency}} (Rio de Janeiro, Brazil) \emph{(\bibinfo{series}{FAccT '24})}. \bibinfo{publisher}{Association for Computing Machinery}, \bibinfo{address}{New York, NY, USA}, \bibinfo{pages}{2322–2347}.
\newblock
\showISBNx{9798400704505}
\href{https://doi.org/10.1145/3630106.3659040}{doi:\nolinkurl{10.1145/3630106.3659040}}


\bibitem[Jensen and Blok(2013)]%
        {jensen2013}
\bibfield{author}{\bibinfo{person}{Casper~Bruun Jensen} {and} \bibinfo{person}{Anders Blok}.} \bibinfo{year}{2013}\natexlab{}.
\newblock \showarticletitle{Techno-animism in Japan: Shinto Cosmograms, Actor-network Theory, and the Enabling Powers of Non-human Agencies}.
\newblock \bibinfo{journal}{\emph{Theory, Culture \& Society}} \bibinfo{volume}{30}, \bibinfo{number}{2} (\bibinfo{year}{2013}), \bibinfo{pages}{84--115}.
\newblock
\showeprint{https://doi.org/10.1177/0263276412456564}
\href{https://doi.org/10.1177/0263276412456564}{doi:\nolinkurl{10.1177/0263276412456564}}


\bibitem[Jensen et~al\mbox{.}(2021)]%
        {jensen2021}
\bibfield{author}{\bibinfo{person}{Theodore Jensen}, \bibinfo{person}{Mohammad Maifi~Hasan Khan}, \bibinfo{person}{Md~Abdullah~Al Fahim}, {and} \bibinfo{person}{Yusuf Albayram}.} \bibinfo{year}{2021}\natexlab{}.
\newblock \showarticletitle{Trust and Anthropomorphism in Tandem: The Interrelated Nature of Automated Agent Appearance and Reliability in Trustworthiness Perceptions}. In \bibinfo{booktitle}{\emph{Proceedings of the 2021 ACM Designing Interactive Systems Conference}} (Virtual Event, USA) \emph{(\bibinfo{series}{DIS '21})}. \bibinfo{publisher}{Association for Computing Machinery}, \bibinfo{address}{New York, NY, USA}, \bibinfo{pages}{1470–1480}.
\newblock
\showISBNx{9781450384766}
\href{https://doi.org/10.1145/3461778.3462102}{doi:\nolinkurl{10.1145/3461778.3462102}}


\bibitem[Kaate et~al\mbox{.}(2024)]%
        {kaate2024}
\bibfield{author}{\bibinfo{person}{Ilkka Kaate}, \bibinfo{person}{Joni Salminen}, \bibinfo{person}{Soon-Gyo Jung}, \bibinfo{person}{Nina Rizun}, \bibinfo{person}{Aleksandra Revina}, {and} \bibinfo{person}{Bernard~J Jansen}.} \bibinfo{year}{2024}\natexlab{}.
\newblock \showarticletitle{Getting Emotional Enough: Analyzing Emotional Diversity in Deepfake Avatars}. In \bibinfo{booktitle}{\emph{Proceedings of the 13th Nordic Conference on Human-Computer Interaction}} (Uppsala, Sweden) \emph{(\bibinfo{series}{NordiCHI '24})}. \bibinfo{publisher}{Association for Computing Machinery}, \bibinfo{address}{New York, NY, USA}, Article \bibinfo{articleno}{62}, \bibinfo{numpages}{12}~pages.
\newblock
\showISBNx{9798400709661}
\href{https://doi.org/10.1145/3679318.3685398}{doi:\nolinkurl{10.1145/3679318.3685398}}


\bibitem[Keysar and Barr(2002)]%
        {keysar2002self}
\bibfield{author}{\bibinfo{person}{Boaz Keysar} {and} \bibinfo{person}{Dale~J Barr}.} \bibinfo{year}{2002}\natexlab{}.
\newblock \showarticletitle{Self-anchoring in conversation: Why language users do not do what they “should.”}.
\newblock \bibinfo{journal}{\emph{Heuristics and biases: The psychology of intuitive judgment}} (\bibinfo{year}{2002}), \bibinfo{pages}{150--166}.
\newblock


\bibitem[Kim et~al\mbox{.}(2022)]%
        {kim2022}
\bibfield{author}{\bibinfo{person}{Tae~Woo Kim}, \bibinfo{person}{Li Jiang}, \bibinfo{person}{Adam Duhachek}, \bibinfo{person}{Hyejin Lee}, {and} \bibinfo{person}{Aaron Garvey}.} \bibinfo{year}{2022}\natexlab{}.
\newblock \showarticletitle{Do You Mind if I Ask You a Personal Question? How AI Service Agents Alter Consumer Self-Disclosure}.
\newblock \bibinfo{journal}{\emph{Journal of Service Research}} \bibinfo{volume}{25}, \bibinfo{number}{4} (\bibinfo{year}{2022}), \bibinfo{pages}{649--666}.
\newblock
\showeprint{https://doi.org/10.1177/10946705221120232}
\href{https://doi.org/10.1177/10946705221120232}{doi:\nolinkurl{10.1177/10946705221120232}}


\bibitem[Knijnenburg and Willemsen(2016)]%
        {knijnenburg2016inferring}
\bibfield{author}{\bibinfo{person}{Bart~P. Knijnenburg} {and} \bibinfo{person}{Martijn~C. Willemsen}.} \bibinfo{year}{2016}\natexlab{}.
\newblock \showarticletitle{Inferring Capabilities of Intelligent Agents from Their External Traits}.
\newblock \bibinfo{journal}{\emph{ACM Trans. Interact. Intell. Syst.}} \bibinfo{volume}{6}, \bibinfo{number}{4}, Article \bibinfo{articleno}{28} (\bibinfo{date}{Nov.} \bibinfo{year}{2016}), \bibinfo{numpages}{25}~pages.
\newblock
\showISSN{2160-6455}
\href{https://doi.org/10.1145/2963106}{doi:\nolinkurl{10.1145/2963106}}


\bibitem[Konya-Baumbach et~al\mbox{.}(2023)]%
        {konya2023someone}
\bibfield{author}{\bibinfo{person}{Elisa Konya-Baumbach}, \bibinfo{person}{Miriam Biller}, {and} \bibinfo{person}{Sergej von Janda}.} \bibinfo{year}{2023}\natexlab{}.
\newblock \showarticletitle{Someone out there? A study on the social presence of anthropomorphized chatbots}.
\newblock \bibinfo{journal}{\emph{Computers in Human Behavior}}  \bibinfo{volume}{139} (\bibinfo{year}{2023}), \bibinfo{pages}{107513}.
\newblock


\bibitem[Krueger(2014)]%
        {krueger2014focus}
\bibfield{author}{\bibinfo{person}{Richard~A Krueger}.} \bibinfo{year}{2014}\natexlab{}.
\newblock \bibinfo{booktitle}{\emph{Focus Groups: A Practical Guide for Applied Research}}.
\newblock \bibinfo{publisher}{Sage Publications}, \bibinfo{address}{Thousand Oaks, CA}.
\newblock


\bibitem[Laestadius et~al\mbox{.}(2024)]%
        {laestadius2024too}
\bibfield{author}{\bibinfo{person}{Linnea Laestadius}, \bibinfo{person}{Andrea Bishop}, \bibinfo{person}{Michael Gonzalez}, \bibinfo{person}{Diana Illen{\v{c}}{\'\i}k}, {and} \bibinfo{person}{Celeste Campos-Castillo}.} \bibinfo{year}{2024}\natexlab{}.
\newblock \showarticletitle{Too human and not human enough: A grounded theory analysis of mental health harms from emotional dependence on the social chatbot Replika}.
\newblock \bibinfo{journal}{\emph{New Media \& Society}} \bibinfo{volume}{26}, \bibinfo{number}{10} (\bibinfo{year}{2024}), \bibinfo{pages}{5923--5941}.
\newblock


\bibitem[Leaver and Srdarov(2023)]%
        {leaver2023chatgpt}
\bibfield{author}{\bibinfo{person}{Tama Leaver} {and} \bibinfo{person}{Suzanne Srdarov}.} \bibinfo{year}{2023}\natexlab{}.
\newblock \showarticletitle{ChatGPT isn't magic: the hype and hypocrisy of generative artificial intelligence (AI) rhetoric}.
\newblock \bibinfo{journal}{\emph{M/c Journal}} \bibinfo{volume}{26}, \bibinfo{number}{5} (\bibinfo{year}{2023}).
\newblock


\bibitem[Lee(2010)]%
        {lee2010more}
\bibfield{author}{\bibinfo{person}{Eun-Ju Lee}.} \bibinfo{year}{2010}\natexlab{}.
\newblock \showarticletitle{The more humanlike, the better? How speech type and users’ cognitive style affect social responses to computers}.
\newblock \bibinfo{journal}{\emph{Computers in Human Behavior}} \bibinfo{volume}{26}, \bibinfo{number}{4} (\bibinfo{year}{2010}), \bibinfo{pages}{665--672}.
\newblock


\bibitem[Leveson(2016)]%
        {leveson2016engineering}
\bibfield{author}{\bibinfo{person}{Nancy~G. Leveson}.} \bibinfo{year}{2016}\natexlab{}.
\newblock \bibinfo{booktitle}{\emph{Engineering a Safer World: Systems Thinking Applied to Safety}}.
\newblock \bibinfo{publisher}{The MIT Press}, \bibinfo{address}{Boston, MA}.
\newblock


\bibitem[Li and Suh(2022)]%
        {li2022anthropomorphism}
\bibfield{author}{\bibinfo{person}{Mengjun Li} {and} \bibinfo{person}{Ayoung Suh}.} \bibinfo{year}{2022}\natexlab{}.
\newblock \showarticletitle{Anthropomorphism in AI-enabled technology: A literature review}.
\newblock \bibinfo{journal}{\emph{Electronic Markets}} \bibinfo{volume}{32}, \bibinfo{number}{4} (\bibinfo{year}{2022}), \bibinfo{pages}{2245--2275}.
\newblock


\bibitem[Li and Sung(2021)]%
        {li2021}
\bibfield{author}{\bibinfo{person}{Xinge Li} {and} \bibinfo{person}{Yongjun Sung}.} \bibinfo{year}{2021}\natexlab{}.
\newblock \showarticletitle{Anthropomorphism brings us closer: The mediating role of psychological distance in User--AI assistant interactions}.
\newblock \bibinfo{journal}{\emph{Computers in human behavior}}  \bibinfo{volume}{118} (\bibinfo{year}{2021}), \bibinfo{pages}{106680}.
\newblock


\bibitem[Liew et~al\mbox{.}(2022)]%
        {liew2022anthropomorphizing}
\bibfield{author}{\bibinfo{person}{Tze~Wei Liew}, \bibinfo{person}{Wei~Ming Pang}, \bibinfo{person}{Meng~Chew Leow}, {and} \bibinfo{person}{Su-Mae Tan}.} \bibinfo{year}{2022}\natexlab{}.
\newblock \showarticletitle{Anthropomorphizing malware, bots, and servers with human-like images and dialogues: The emotional design effects in a multimedia learning environment}.
\newblock \bibinfo{journal}{\emph{Smart Learning Environments}} \bibinfo{volume}{9}, \bibinfo{number}{1} (\bibinfo{year}{2022}), \bibinfo{pages}{5}.
\newblock


\bibitem[Lipton and Steinhardt(2019)]%
        {lipton2019}
\bibfield{author}{\bibinfo{person}{Zachary~C. Lipton} {and} \bibinfo{person}{Jacob Steinhardt}.} \bibinfo{year}{2019}\natexlab{}.
\newblock \showarticletitle{Troubling Trends in Machine Learning Scholarship: Some ML papers suffer from flaws that could mislead the public and stymie future research.}
\newblock \bibinfo{journal}{\emph{Queue}} \bibinfo{volume}{17}, \bibinfo{number}{1} (\bibinfo{date}{Feb.} \bibinfo{year}{2019}), \bibinfo{pages}{45–77}.
\newblock
\showISSN{1542-7730}
\href{https://doi.org/10.1145/3317287.3328534}{doi:\nolinkurl{10.1145/3317287.3328534}}


\bibitem[Ljungblad et~al\mbox{.}(2012)]%
        {ljungblad2012hospital}
\bibfield{author}{\bibinfo{person}{Sara Ljungblad}, \bibinfo{person}{Jirina Kotrbova}, \bibinfo{person}{Mattias Jacobsson}, \bibinfo{person}{Henriette Cramer}, {and} \bibinfo{person}{Karol Niechwiadowicz}.} \bibinfo{year}{2012}\natexlab{}.
\newblock \showarticletitle{Hospital robot at work: something alien or an intelligent colleague?}. In \bibinfo{booktitle}{\emph{Proceedings of the ACM 2012 Conference on Computer Supported Cooperative Work}} (Seattle, Washington, USA) \emph{(\bibinfo{series}{CSCW '12})}. \bibinfo{publisher}{Association for Computing Machinery}, \bibinfo{address}{New York, NY, USA}, \bibinfo{pages}{177–186}.
\newblock
\showISBNx{9781450310864}
\href{https://doi.org/10.1145/2145204.2145233}{doi:\nolinkurl{10.1145/2145204.2145233}}


\bibitem[Luka(2024)]%
        {replika2024}
\bibfield{author}{\bibinfo{person}{Inc Luka}.} \bibinfo{year}{2024}\natexlab{}.
\newblock \bibinfo{title}{Replika}.
\newblock
\urldef\tempurl%
\url{https://replika.com/}
\showURL{%
\tempurl}


\bibitem[Lupton and Noble(1997)]%
        {lupton1997}
\bibfield{author}{\bibinfo{person}{Debora Lupton} {and} \bibinfo{person}{Greg Noble}.} \bibinfo{year}{1997}\natexlab{}.
\newblock \showarticletitle{Just a Machine? Dehumanizing Strategies in Personal Computer Use}.
\newblock \bibinfo{journal}{\emph{Body \& Society}} \bibinfo{volume}{3}, \bibinfo{number}{2} (\bibinfo{year}{1997}), \bibinfo{pages}{83--101}.
\newblock
\showeprint{https://doi.org/10.1177/1357034X97003002006}
\href{https://doi.org/10.1177/1357034X97003002006}{doi:\nolinkurl{10.1177/1357034X97003002006}}


\bibitem[Madaio et~al\mbox{.}(2024)]%
        {madaio2024learning}
\bibfield{author}{\bibinfo{person}{Michael Madaio}, \bibinfo{person}{Shivani Kapania}, \bibinfo{person}{Rida Qadri}, \bibinfo{person}{Ding Wang}, \bibinfo{person}{Andrew Zaldivar}, \bibinfo{person}{Remi Denton}, {and} \bibinfo{person}{Lauren Wilcox}.} \bibinfo{year}{2024}\natexlab{}.
\newblock \showarticletitle{Learning about responsible AI on-the-job: Learning pathways, orientations, and aspirations}. In \bibinfo{booktitle}{\emph{Proceedings of the 2024 ACM Conference on Fairness, Accountability, and Transparency}}. \bibinfo{pages}{1544--1558}.
\newblock


\bibitem[Maeda and Quan-Haase(2024)]%
        {maeda2024}
\bibfield{author}{\bibinfo{person}{Takuya Maeda} {and} \bibinfo{person}{Anabel Quan-Haase}.} \bibinfo{year}{2024}\natexlab{}.
\newblock \showarticletitle{When Human-AI Interactions Become Parasocial: Agency and Anthropomorphism in Affective Design}. In \bibinfo{booktitle}{\emph{Proceedings of the 2024 ACM Conference on Fairness, Accountability, and Transparency}} (Rio de Janeiro, Brazil) \emph{(\bibinfo{series}{FAccT '24})}. \bibinfo{publisher}{Association for Computing Machinery}, \bibinfo{address}{New York, NY, USA}, \bibinfo{pages}{1068–1077}.
\newblock
\showISBNx{9798400704505}
\href{https://doi.org/10.1145/3630106.3658956}{doi:\nolinkurl{10.1145/3630106.3658956}}


\bibitem[Manzini et~al\mbox{.}(2024)]%
        {manzini2024should}
\bibfield{author}{\bibinfo{person}{Arianna Manzini}, \bibinfo{person}{Geoff Keeling}, \bibinfo{person}{Nahema Marchal}, \bibinfo{person}{Kevin~R. McKee}, \bibinfo{person}{Verena Rieser}, {and} \bibinfo{person}{Iason Gabriel}.} \bibinfo{year}{2024}\natexlab{}.
\newblock \showarticletitle{Should Users Trust Advanced AI Assistants? Justified Trust As a Function of Competence and Alignment}. In \bibinfo{booktitle}{\emph{Proceedings of the 2024 ACM Conference on Fairness, Accountability, and Transparency}} (Rio de Janeiro, Brazil) \emph{(\bibinfo{series}{FAccT '24})}. \bibinfo{publisher}{Association for Computing Machinery}, \bibinfo{address}{New York, NY, USA}, \bibinfo{pages}{1174–1186}.
\newblock
\showISBNx{9798400704505}
\href{https://doi.org/10.1145/3630106.3658964}{doi:\nolinkurl{10.1145/3630106.3658964}}


\bibitem[Maris et~al\mbox{.}(2022)]%
        {maris2022}
\bibfield{author}{\bibinfo{person}{Elena Maris}, \bibinfo{person}{Kelly~B. Wagman}, \bibinfo{person}{Rachel Bergmann}, {and} \bibinfo{person}{Danielle Bragg}.} \bibinfo{year}{2022}\natexlab{}.
\newblock \showarticletitle{Tech Worker Perspectives on Considering the Interpersonal Implications of Communication Technologies}.
\newblock \bibinfo{journal}{\emph{Proc. ACM Hum.-Comput. Interact.}} \bibinfo{volume}{7}, \bibinfo{number}{GROUP}, Article \bibinfo{articleno}{16} (\bibinfo{date}{Dec.} \bibinfo{year}{2022}), \bibinfo{numpages}{22}~pages.
\newblock
\href{https://doi.org/10.1145/3567566}{doi:\nolinkurl{10.1145/3567566}}


\bibitem[Masrani et~al\mbox{.}(2023)]%
        {masrani2023}
\bibfield{author}{\bibinfo{person}{Teale~W. Masrani}, \bibinfo{person}{Jack Jamieson}, \bibinfo{person}{Naomi Yamashita}, {and} \bibinfo{person}{Helen~Ai He}.} \bibinfo{year}{2023}\natexlab{}.
\newblock \showarticletitle{Slowing it Down: Towards Facilitating Interpersonal Mindfulness in Online Polarizing Conversations Over Social Media}.
\newblock \bibinfo{journal}{\emph{Proc. ACM Hum.-Comput. Interact.}} \bibinfo{volume}{7}, \bibinfo{number}{CSCW1}, Article \bibinfo{articleno}{90} (\bibinfo{date}{April} \bibinfo{year}{2023}), \bibinfo{numpages}{27}~pages.
\newblock
\href{https://doi.org/10.1145/3579523}{doi:\nolinkurl{10.1145/3579523}}


\bibitem[Mehrotra et~al\mbox{.}(2024)]%
        {mehrotra2024}
\bibfield{author}{\bibinfo{person}{Siddharth Mehrotra}, \bibinfo{person}{Chadha Degachi}, \bibinfo{person}{Oleksandra Vereschak}, \bibinfo{person}{Catholijn~M. Jonker}, {and} \bibinfo{person}{Myrthe~L. Tielman}.} \bibinfo{year}{2024}\natexlab{}.
\newblock \bibinfo{title}{A Systematic Review on Fostering Appropriate Trust in Human-AI Interaction: Trends, Opportunities and Challenges}.
\newblock
\href{https://doi.org/10.1145/3696449}{doi:\nolinkurl{10.1145/3696449}}
\newblock
\shownote{Just Accepted}.


\bibitem[Miesler(2012)]%
        {miesler2012product}
\bibfield{author}{\bibinfo{person}{Linda Miesler}.} \bibinfo{year}{2012}\natexlab{}.
\newblock \showarticletitle{Product choice and anthropomorphic designs: Do consumption goals shape innate preferences for human-like forms?}
\newblock \bibinfo{journal}{\emph{The Design Journal}} \bibinfo{volume}{15}, \bibinfo{number}{3} (\bibinfo{year}{2012}), \bibinfo{pages}{373--392}.
\newblock


\bibitem[Minsky(2007)]%
        {minsky2007emotion}
\bibfield{author}{\bibinfo{person}{Marvin Minsky}.} \bibinfo{year}{2007}\natexlab{}.
\newblock \bibinfo{booktitle}{\emph{The Emotion Machine: Commonsense Thinking, Artificial Intelligence, and the Future of the Human Mind}}.
\newblock \bibinfo{publisher}{Simon and Schuster}, \bibinfo{address}{New York, New York, USA}.
\newblock


\bibitem[Mitchell et~al\mbox{.}(1997)]%
        {mitchell1997anthropomorphism}
\bibfield{author}{\bibinfo{person}{Robert~W. Mitchell}, \bibinfo{person}{Nicholas~S. Thompson}, {and} \bibinfo{person}{H.~Lyn Miles}.} \bibinfo{year}{1997}\natexlab{}.
\newblock \bibinfo{booktitle}{\emph{Anthropomorphism, Anecdotes, and Animals}}.
\newblock \bibinfo{publisher}{SUNY Press}, \bibinfo{address}{Albany, NY, USA}.
\newblock


\bibitem[Moon(2000)]%
        {moon2000intimate}
\bibfield{author}{\bibinfo{person}{Youngme Moon}.} \bibinfo{year}{2000}\natexlab{}.
\newblock \showarticletitle{Intimate exchanges: Using computers to elicit self-disclosure from consumers}.
\newblock \bibinfo{journal}{\emph{Journal of consumer research}} \bibinfo{volume}{26}, \bibinfo{number}{4} (\bibinfo{year}{2000}), \bibinfo{pages}{323--339}.
\newblock


\bibitem[Moon and Nass(1996)]%
        {moon1996}
\bibfield{author}{\bibinfo{person}{Youngme Moon} {and} \bibinfo{person}{Clifford Nass}.} \bibinfo{year}{1996}\natexlab{}.
\newblock \showarticletitle{How “Real” Are Computer Personalities?: Psychological Responses to Personality Types in Human-Computer Interaction}.
\newblock \bibinfo{journal}{\emph{Communication Research}} \bibinfo{volume}{23}, \bibinfo{number}{6} (\bibinfo{year}{1996}), \bibinfo{pages}{651--674}.
\newblock
\showeprint{https://doi.org/10.1177/009365096023006002}
\href{https://doi.org/10.1177/009365096023006002}{doi:\nolinkurl{10.1177/009365096023006002}}


\bibitem[Morgan(1996)]%
        {morgan1996focus}
\bibfield{author}{\bibinfo{person}{David~L. Morgan}.} \bibinfo{year}{1996}\natexlab{}.
\newblock \bibinfo{booktitle}{\emph{Focus Groups as Qualitative Research}}.
\newblock \bibinfo{publisher}{Sage publications}, \bibinfo{address}{Thousand Oaks, CA}.
\newblock


\bibitem[Morris et~al\mbox{.}(2007)]%
        {morris2007metaphors}
\bibfield{author}{\bibinfo{person}{Michael~W Morris}, \bibinfo{person}{Oliver~J Sheldon}, \bibinfo{person}{Daniel~R Ames}, {and} \bibinfo{person}{Maia~J Young}.} \bibinfo{year}{2007}\natexlab{}.
\newblock \showarticletitle{Metaphors and the market: Consequences and preconditions of agent and object metaphors in stock market commentary}.
\newblock \bibinfo{journal}{\emph{Organizational behavior and human decision processes}} \bibinfo{volume}{102}, \bibinfo{number}{2} (\bibinfo{year}{2007}), \bibinfo{pages}{174--192}.
\newblock


\bibitem[Moussawi and Benbunan-Fich(2021)]%
        {moussawi2021effect}
\bibfield{author}{\bibinfo{person}{Sara Moussawi} {and} \bibinfo{person}{Raquel Benbunan-Fich}.} \bibinfo{year}{2021}\natexlab{}.
\newblock \showarticletitle{The effect of voice and humour on users’ perceptions of personal intelligent agents}.
\newblock \bibinfo{journal}{\emph{Behaviour \& Information Technology}} \bibinfo{volume}{40}, \bibinfo{number}{15} (\bibinfo{year}{2021}), \bibinfo{pages}{1603--1626}.
\newblock


\bibitem[Muller(2004)]%
        {muller2004}
\bibfield{author}{\bibinfo{person}{Michael Muller}.} \bibinfo{year}{2004}\natexlab{}.
\newblock \showarticletitle{Multiple paradigms in affective computing}.
\newblock \bibinfo{journal}{\emph{Interacting with Computers}} \bibinfo{volume}{16}, \bibinfo{number}{4} (\bibinfo{year}{2004}), \bibinfo{pages}{759--768}.
\newblock
\href{https://doi.org/10.1016/j.intcom.2004.06.005}{doi:\nolinkurl{10.1016/j.intcom.2004.06.005}}


\bibitem[Nass et~al\mbox{.}(1996)]%
        {nass1996can}
\bibfield{author}{\bibinfo{person}{Clifford Nass}, \bibinfo{person}{Brian~Jeffrey Fogg}, {and} \bibinfo{person}{Youngme Moon}.} \bibinfo{year}{1996}\natexlab{}.
\newblock \showarticletitle{Can computers be teammates?}
\newblock \bibinfo{journal}{\emph{International Journal of Human-Computer Studies}} \bibinfo{volume}{45}, \bibinfo{number}{6} (\bibinfo{year}{1996}), \bibinfo{pages}{669--678}.
\newblock


\bibitem[Nass and Moon(2000)]%
        {nass2000machines}
\bibfield{author}{\bibinfo{person}{Clifford Nass} {and} \bibinfo{person}{Youngme Moon}.} \bibinfo{year}{2000}\natexlab{}.
\newblock \showarticletitle{Machines and mindlessness: Social responses to computers}.
\newblock \bibinfo{journal}{\emph{Journal of social issues}} \bibinfo{volume}{56}, \bibinfo{number}{1} (\bibinfo{year}{2000}), \bibinfo{pages}{81--103}.
\newblock


\bibitem[Nass et~al\mbox{.}(1997)]%
        {nass1997machines}
\bibfield{author}{\bibinfo{person}{Clifford Nass}, \bibinfo{person}{Youngme Moon}, {and} \bibinfo{person}{Nancy Green}.} \bibinfo{year}{1997}\natexlab{}.
\newblock \showarticletitle{Are machines gender neutral? Gender-stereotypic responses to computers with voices}.
\newblock \bibinfo{journal}{\emph{Journal of applied social psychology}} \bibinfo{volume}{27}, \bibinfo{number}{10} (\bibinfo{year}{1997}), \bibinfo{pages}{864--876}.
\newblock


\bibitem[Nass et~al\mbox{.}(1993)]%
        {nass1993anthropomorphism}
\bibfield{author}{\bibinfo{person}{Clifford Nass}, \bibinfo{person}{Jonathan Steuer}, \bibinfo{person}{Ellen Tauber}, {and} \bibinfo{person}{Heidi Reeder}.} \bibinfo{year}{1993}\natexlab{}.
\newblock \bibinfo{title}{Anthropomorphism, agency, and ethopoeia: computers as social actors}.
\newblock \bibinfo{numpages}{111--112}~pages.
\newblock


\bibitem[Nelkin(1992)]%
        {nelkin1992controversy}
\bibfield{author}{\bibinfo{person}{Dorothy~Ed Nelkin}.} \bibinfo{year}{1992}\natexlab{}.
\newblock \bibinfo{booktitle}{\emph{Controversy: Politics of Technical Decisions}}.
\newblock \bibinfo{publisher}{Sage Publications, Inc}, \bibinfo{address}{Thousand Oaks, CA}.
\newblock


\bibitem[Nickerson(1999)]%
        {nickerson1999we}
\bibfield{author}{\bibinfo{person}{Raymond~S Nickerson}.} \bibinfo{year}{1999}\natexlab{}.
\newblock \showarticletitle{How we know—and sometimes misjudge—what others know: Imputing one's own knowledge to others.}
\newblock \bibinfo{journal}{\emph{Psychological bulletin}} \bibinfo{volume}{125}, \bibinfo{number}{6} (\bibinfo{year}{1999}), \bibinfo{pages}{737}.
\newblock


\bibitem[on~AI~Governance(2023)]%
        {oecd2022}
\bibfield{author}{\bibinfo{person}{OECD Working~Party on AI~Governance}.} \bibinfo{year}{2023}\natexlab{}.
\newblock \bibinfo{title}{Summary of OECD expert discussion on future risks from artificial intelligence of 20 October 2022}.
\newblock \bibinfo{howpublished}{\url{https://wp.oecd.ai/app/uploads/2023/03/OECD-Foresight-workshop-notes-1.pdf}}.
\newblock


\bibitem[Palinkas et~al\mbox{.}(2015)]%
        {palinkas2015purposeful}
\bibfield{author}{\bibinfo{person}{Lawrence~A Palinkas}, \bibinfo{person}{Sarah~M Horwitz}, \bibinfo{person}{Carla~A Green}, \bibinfo{person}{Jennifer~P Wisdom}, \bibinfo{person}{Naihua Duan}, {and} \bibinfo{person}{Kimberly Hoagwood}.} \bibinfo{year}{2015}\natexlab{}.
\newblock \showarticletitle{Purposeful sampling for qualitative data collection and analysis in mixed method implementation research}.
\newblock \bibinfo{journal}{\emph{Administration and policy in mental health and mental health services research}}  \bibinfo{volume}{42} (\bibinfo{year}{2015}), \bibinfo{pages}{533--544}.
\newblock


\bibitem[Pawlik(2021)]%
        {pawlik2021design}
\bibfield{author}{\bibinfo{person}{V~Phoebe Pawlik}.} \bibinfo{year}{2021}\natexlab{}.
\newblock \bibinfo{title}{Design matters! How visual gendered anthropomorphic design cues moderate the determinants of the behavioral intention towards using chatbots}.
\newblock \bibinfo{numpages}{192--208}~pages.
\newblock


\bibitem[Pelau et~al\mbox{.}(2021)]%
        {PELAU2021106855}
\bibfield{author}{\bibinfo{person}{Corina Pelau}, \bibinfo{person}{Dan-Cristian Dabija}, {and} \bibinfo{person}{Irina Ene}.} \bibinfo{year}{2021}\natexlab{}.
\newblock \showarticletitle{What makes an AI device human-like? The role of interaction quality, empathy and perceived psychological anthropomorphic characteristics in the acceptance of artificial intelligence in the service industry}.
\newblock \bibinfo{journal}{\emph{Computers in Human Behavior}}  \bibinfo{volume}{122} (\bibinfo{year}{2021}), \bibinfo{pages}{106855}.
\newblock
\showISSN{0747-5632}
\href{https://doi.org/10.1016/j.chb.2021.106855}{doi:\nolinkurl{10.1016/j.chb.2021.106855}}


\bibitem[Pentina et~al\mbox{.}(2023)]%
        {pentina2023}
\bibfield{author}{\bibinfo{person}{Iryna Pentina}, \bibinfo{person}{Tyler Hancock}, {and} \bibinfo{person}{Tianling Xie}.} \bibinfo{year}{2023}\natexlab{}.
\newblock \showarticletitle{Exploring relationship development with social chatbots: A mixed-method study of replika}.
\newblock \bibinfo{journal}{\emph{Computers in Human Behavior}}  \bibinfo{volume}{140} (\bibinfo{year}{2023}), \bibinfo{pages}{107600}.
\newblock
\showISSN{0747-5632}
\href{https://doi.org/10.1016/j.chb.2022.107600}{doi:\nolinkurl{10.1016/j.chb.2022.107600}}


\bibitem[Pfeuffer et~al\mbox{.}(2019)]%
        {pfeuffer2019}
\bibfield{author}{\bibinfo{person}{Nicolas Pfeuffer}, \bibinfo{person}{Alexander Benlian}, \bibinfo{person}{Henner Gimpel}, {and} \bibinfo{person}{Oliver Hinz}.} \bibinfo{year}{2019}\natexlab{}.
\newblock \showarticletitle{Anthropomorphic information systems}.
\newblock \bibinfo{journal}{\emph{Business \& Information Systems Engineering}}  \bibinfo{volume}{61} (\bibinfo{year}{2019}), \bibinfo{pages}{523--533}.
\newblock


\bibitem[Placani(2024)]%
        {placani2024anthropomorphism}
\bibfield{author}{\bibinfo{person}{Adriana Placani}.} \bibinfo{year}{2024}\natexlab{}.
\newblock \showarticletitle{Anthropomorphism in AI: hype and fallacy}.
\newblock \bibinfo{journal}{\emph{AI and Ethics}}  \bibinfo{volume}{4} (\bibinfo{year}{2024}), \bibinfo{pages}{1--8}.
\newblock


\bibitem[Qiu and Benbasat(2009)]%
        {qiu2009evaluating}
\bibfield{author}{\bibinfo{person}{Lingyun Qiu} {and} \bibinfo{person}{Izak Benbasat}.} \bibinfo{year}{2009}\natexlab{}.
\newblock \showarticletitle{Evaluating anthropomorphic product recommendation agents: A social relationship perspective to designing information systems}.
\newblock \bibinfo{journal}{\emph{Journal of management information systems}} \bibinfo{volume}{25}, \bibinfo{number}{4} (\bibinfo{year}{2009}), \bibinfo{pages}{145--182}.
\newblock


\bibitem[Quang-An~Ha and Capistrano(2021)]%
        {ha2021}
\bibfield{author}{\bibinfo{person}{Ha~Uy~Uy Quang-An~Ha, Jengchung Victor~Chen} {and} \bibinfo{person}{Erik~Paolo Capistrano}.} \bibinfo{year}{2021}\natexlab{}.
\newblock \showarticletitle{Exploring the Privacy Concerns in Using Intelligent Virtual Assistants under Perspectives of Information Sensitivity and Anthropomorphism}.
\newblock \bibinfo{journal}{\emph{International Journal of Human–Computer Interaction}} \bibinfo{volume}{37}, \bibinfo{number}{6} (\bibinfo{year}{2021}), \bibinfo{pages}{512--527}.
\newblock
\showeprint{https://doi.org/10.1080/10447318.2020.1834728}
\href{https://doi.org/10.1080/10447318.2020.1834728}{doi:\nolinkurl{10.1080/10447318.2020.1834728}}


\bibitem[Rajaobelina et~al\mbox{.}(2021)]%
        {rajaobelina2021creepiness}
\bibfield{author}{\bibinfo{person}{Lova Rajaobelina}, \bibinfo{person}{Sandrine Prom~Tep}, \bibinfo{person}{Manon Arcand}, {and} \bibinfo{person}{Line Ricard}.} \bibinfo{year}{2021}\natexlab{}.
\newblock \showarticletitle{Creepiness: Its antecedents and impact on loyalty when interacting with a chatbot}.
\newblock \bibinfo{journal}{\emph{Psychology \& Marketing}} \bibinfo{volume}{38}, \bibinfo{number}{12} (\bibinfo{year}{2021}), \bibinfo{pages}{2339--2356}.
\newblock


\bibitem[Rakova et~al\mbox{.}(2021)]%
        {rakova2021responsible}
\bibfield{author}{\bibinfo{person}{Bogdana Rakova}, \bibinfo{person}{Jingying Yang}, \bibinfo{person}{Henriette Cramer}, {and} \bibinfo{person}{Rumman Chowdhury}.} \bibinfo{year}{2021}\natexlab{}.
\newblock \showarticletitle{Where responsible AI meets reality: Practitioner perspectives on enablers for shifting organizational practices}.
\newblock \bibinfo{journal}{\emph{Proceedings of the ACM on Human-Computer Interaction}} \bibinfo{volume}{5}, \bibinfo{number}{CSCW1} (\bibinfo{year}{2021}), \bibinfo{pages}{1--23}.
\newblock


\bibitem[Rasmussen(1994)]%
        {rasmussen1994risk}
\bibfield{author}{\bibinfo{person}{Jens Rasmussen}.} \bibinfo{year}{1994}\natexlab{}.
\newblock \showarticletitle{Risk management, adaptation, and design for safety}.
\newblock In \bibinfo{booktitle}{\emph{Future Risks and Risk Management}}. \bibinfo{publisher}{Springer}, \bibinfo{address}{Dordrecht, Netherlands}, \bibinfo{pages}{1--36}.
\newblock


\bibitem[Reeves and Nass(1996)]%
        {reeves1996media}
\bibfield{author}{\bibinfo{person}{Byron Reeves} {and} \bibinfo{person}{Clifford Nass}.} \bibinfo{year}{1996}\natexlab{}.
\newblock \showarticletitle{The media equation: How people treat computers, television, and new media like real people}.
\newblock \bibinfo{journal}{\emph{Cambridge, UK}} \bibinfo{volume}{10}, \bibinfo{number}{10} (\bibinfo{year}{1996}), \bibinfo{pages}{19--36}.
\newblock


\bibitem[Research(2024)]%
        {chai2024}
\bibfield{author}{\bibinfo{person}{Chai Research}.} \bibinfo{year}{2024}\natexlab{}.
\newblock \bibinfo{title}{CHAI}.
\newblock
\urldef\tempurl%
\url{https://www.chai-research.com/}
\showURL{%
\tempurl}


\bibitem[Riedl et~al\mbox{.}(2011)]%
        {riedl2011trusting}
\bibfield{author}{\bibinfo{person}{Ren{\'e} Riedl}, \bibinfo{person}{Peter Mohr}, \bibinfo{person}{Peter Kenning}, \bibinfo{person}{Fred Davis}, {and} \bibinfo{person}{Hauke Heekeren}.} \bibinfo{year}{2011}\natexlab{}.
\newblock \bibinfo{title}{Trusting humans and avatars: Behavioral and neural evidence}.
\newblock


\bibitem[Rip and Kemp(1998)]%
        {rip1998technological}
\bibfield{author}{\bibinfo{person}{Arie Rip} {and} \bibinfo{person}{Ren{\'e} Kemp}.} \bibinfo{year}{1998}\natexlab{}.
\newblock \showarticletitle{Technological change}.
\newblock In \bibinfo{booktitle}{\emph{Human choice and climate change: Vol. II, Resources and Technology}}. \bibinfo{publisher}{Battelle Press}, \bibinfo{address}{Columbus, Ohio}, \bibinfo{pages}{327--399}.
\newblock


\bibitem[Rismani et~al\mbox{.}(2023)]%
        {rismani2023}
\bibfield{author}{\bibinfo{person}{Shalaleh Rismani}, \bibinfo{person}{Renee Shelby}, \bibinfo{person}{Andrew Smart}, \bibinfo{person}{Edgar Jatho}, \bibinfo{person}{Joshua Kroll}, \bibinfo{person}{AJung Moon}, {and} \bibinfo{person}{Negar Rostamzadeh}.} \bibinfo{year}{2023}\natexlab{}.
\newblock \showarticletitle{From Plane Crashes to Algorithmic Harm: Applicability of Safety Engineering Frameworks for Responsible ML}. In \bibinfo{booktitle}{\emph{Proceedings of the 2023 CHI Conference on Human Factors in Computing Systems}} (Hamburg, Germany) \emph{(\bibinfo{series}{CHI '23})}. \bibinfo{publisher}{Association for Computing Machinery}, \bibinfo{address}{New York, NY, USA}, Article \bibinfo{articleno}{2}, \bibinfo{numpages}{18}~pages.
\newblock
\showISBNx{9781450394215}
\href{https://doi.org/10.1145/3544548.3581407}{doi:\nolinkurl{10.1145/3544548.3581407}}


\bibitem[Roberts et~al\mbox{.}(2019)]%
        {roberts2019attempting}
\bibfield{author}{\bibinfo{person}{Kate Roberts}, \bibinfo{person}{Anthony Dowell}, {and} \bibinfo{person}{Jing-Bao Nie}.} \bibinfo{year}{2019}\natexlab{}.
\newblock \showarticletitle{Attempting rigour and replicability in thematic analysis of qualitative research data; a case study of codebook development}.
\newblock \bibinfo{journal}{\emph{BMC Medical Research Methodology}}  \bibinfo{volume}{19} (\bibinfo{year}{2019}), \bibinfo{pages}{1--8}.
\newblock


\bibitem[Sah and Peng(2015)]%
        {sah2015effects}
\bibfield{author}{\bibinfo{person}{Young~June Sah} {and} \bibinfo{person}{Wei Peng}.} \bibinfo{year}{2015}\natexlab{}.
\newblock \showarticletitle{Effects of visual and linguistic anthropomorphic cues on social perception, self-awareness, and information disclosure in a health website}.
\newblock \bibinfo{journal}{\emph{Computers in Human Behavior}}  \bibinfo{volume}{45} (\bibinfo{year}{2015}), \bibinfo{pages}{392--401}.
\newblock


\bibitem[Scheuerman and Brubaker(2024)]%
        {scheuerman2024products}
\bibfield{author}{\bibinfo{person}{Morgan~Klaus Scheuerman} {and} \bibinfo{person}{Jed~R. Brubaker}.} \bibinfo{year}{2024}\natexlab{}.
\newblock \showarticletitle{Products of Positionality: How Tech Workers Shape Identity Concepts in Computer Vision}. In \bibinfo{booktitle}{\emph{Proceedings of the 2024 CHI Conference on Human Factors in Computing Systems}} (Honolulu, HI, USA) \emph{(\bibinfo{series}{CHI '24})}. \bibinfo{publisher}{Association for Computing Machinery}, \bibinfo{address}{New York, NY, USA}, Article \bibinfo{articleno}{762}, \bibinfo{numpages}{18}~pages.
\newblock
\showISBNx{9798400703300}
\href{https://doi.org/10.1145/3613904.3641890}{doi:\nolinkurl{10.1145/3613904.3641890}}


\bibitem[Schiff et~al\mbox{.}(2020)]%
        {schiff2020principles}
\bibfield{author}{\bibinfo{person}{Daniel Schiff}, \bibinfo{person}{Bogdana Rakova}, \bibinfo{person}{Aladdin Ayesh}, \bibinfo{person}{Anat Fanti}, {and} \bibinfo{person}{Michael Lennon}.} \bibinfo{year}{2020}\natexlab{}.
\newblock \bibinfo{title}{Principles to Practices for Responsible AI: Closing the Gap}.
\newblock
\showeprint[arxiv]{2006.04707}~[cs.CY]
\urldef\tempurl%
\url{https://arxiv.org/abs/2006.04707}
\showURL{%
\tempurl}


\bibitem[Schneiders et~al\mbox{.}(2022)]%
        {schneiders2021effect}
\bibfield{author}{\bibinfo{person}{Eike Schneiders}, \bibinfo{person}{Eleftherios Papachristos}, {and} \bibinfo{person}{Niels van Berkel}.} \bibinfo{year}{2022}\natexlab{}.
\newblock \showarticletitle{The Effect of Embodied Anthropomorphism of Personal Assistants on User Perceptions}. In \bibinfo{booktitle}{\emph{Proceedings of the 33rd Australian Conference on Human-Computer Interaction}} (Melbourne, VIC, Australia) \emph{(\bibinfo{series}{OzCHI '21})}. \bibinfo{publisher}{Association for Computing Machinery}, \bibinfo{address}{New York, NY, USA}, \bibinfo{pages}{231–241}.
\newblock
\showISBNx{9781450395984}
\href{https://doi.org/10.1145/3520495.3520503}{doi:\nolinkurl{10.1145/3520495.3520503}}


\bibitem[Schuetzler et~al\mbox{.}(2014)]%
        {schuetzler2014facilitating}
\bibfield{author}{\bibinfo{person}{Ryan~M Schuetzler}, \bibinfo{person}{Mark Grimes}, \bibinfo{person}{Justin~Scott Giboney}, {and} \bibinfo{person}{Joesph Buckman}.} \bibinfo{year}{2014}\natexlab{}.
\newblock \bibinfo{title}{Facilitating natural conversational agent interactions: lessons from a deception experiment}.
\newblock


\bibitem[Seeger et~al\mbox{.}(2021)]%
        {seeger2021texting}
\bibfield{author}{\bibinfo{person}{Anna-Maria Seeger}, \bibinfo{person}{Jella Pfeiffer}, {and} \bibinfo{person}{Armin Heinzl}.} \bibinfo{year}{2021}\natexlab{}.
\newblock \showarticletitle{Texting with humanlike conversational agents: Designing for anthropomorphism}.
\newblock \bibinfo{journal}{\emph{Journal of the Association for Information systems}} \bibinfo{volume}{22}, \bibinfo{number}{4} (\bibinfo{year}{2021}), \bibinfo{pages}{8}.
\newblock


\bibitem[Seymour and Van~Kleek(2021)]%
        {seymour2021}
\bibfield{author}{\bibinfo{person}{William Seymour} {and} \bibinfo{person}{Max Van~Kleek}.} \bibinfo{year}{2021}\natexlab{}.
\newblock \showarticletitle{Exploring Interactions Between Trust, Anthropomorphism, and Relationship Development in Voice Assistants}.
\newblock \bibinfo{journal}{\emph{Proc. ACM Hum.-Comput. Interact.}} \bibinfo{volume}{5}, \bibinfo{number}{CSCW2}, Article \bibinfo{articleno}{371} (\bibinfo{date}{Oct.} \bibinfo{year}{2021}), \bibinfo{numpages}{16}~pages.
\newblock
\href{https://doi.org/10.1145/3479515}{doi:\nolinkurl{10.1145/3479515}}


\bibitem[Shanahan(2024)]%
        {shanahan2024talking}
\bibfield{author}{\bibinfo{person}{Murray Shanahan}.} \bibinfo{year}{2024}\natexlab{}.
\newblock \showarticletitle{Talking about large language models}.
\newblock \bibinfo{journal}{\emph{Commun. ACM}} \bibinfo{volume}{67}, \bibinfo{number}{2} (\bibinfo{year}{2024}), \bibinfo{pages}{68--79}.
\newblock


\bibitem[Sheehan et~al\mbox{.}(2020)]%
        {sheehan2020customer}
\bibfield{author}{\bibinfo{person}{Ben Sheehan}, \bibinfo{person}{Hyun~Seung Jin}, {and} \bibinfo{person}{Udo Gottlieb}.} \bibinfo{year}{2020}\natexlab{}.
\newblock \showarticletitle{Customer service chatbots: Anthropomorphism and adoption}.
\newblock \bibinfo{journal}{\emph{Journal of Business Research}}  \bibinfo{volume}{115} (\bibinfo{year}{2020}), \bibinfo{pages}{14--24}.
\newblock


\bibitem[Shelby et~al\mbox{.}(2023)]%
        {shelby2023}
\bibfield{author}{\bibinfo{person}{Renee Shelby}, \bibinfo{person}{Shalaleh Rismani}, \bibinfo{person}{Kathryn Henne}, \bibinfo{person}{AJung Moon}, \bibinfo{person}{Negar Rostamzadeh}, \bibinfo{person}{Paul Nicholas}, \bibinfo{person}{N'Mah Yilla-Akbari}, \bibinfo{person}{Jess Gallegos}, \bibinfo{person}{Andrew Smart}, \bibinfo{person}{Emilio Garcia}, {and} \bibinfo{person}{Gurleen Virk}.} \bibinfo{year}{2023}\natexlab{}.
\newblock \showarticletitle{Sociotechnical Harms of Algorithmic Systems: Scoping a Taxonomy for Harm Reduction}. In \bibinfo{booktitle}{\emph{Proceedings of the 2023 AAAI/ACM Conference on AI, Ethics, and Society}} (Montr\'{e}al, QC, Canada) \emph{(\bibinfo{series}{AIES '23})}. \bibinfo{publisher}{Association for Computing Machinery}, \bibinfo{address}{New York, NY, USA}, \bibinfo{pages}{723–741}.
\newblock
\showISBNx{9798400702310}
\href{https://doi.org/10.1145/3600211.3604673}{doi:\nolinkurl{10.1145/3600211.3604673}}


\bibitem[Shelby et~al\mbox{.}(2024)]%
        {shelby2024}
\bibfield{author}{\bibinfo{person}{Renee Shelby}, \bibinfo{person}{Shalaleh Rismani}, {and} \bibinfo{person}{Negar Rostamzadeh}.} \bibinfo{year}{2024}\natexlab{}.
\newblock \showarticletitle{Generative AI in Creative Practice: ML-Artist Folk Theories of T2I Use, Harm, and Harm-Reduction}. In \bibinfo{booktitle}{\emph{Proceedings of the 2024 CHI Conference on Human Factors in Computing Systems}} (Honolulu, HI, USA) \emph{(\bibinfo{series}{CHI '24})}. \bibinfo{publisher}{Association for Computing Machinery}, \bibinfo{address}{New York, NY, USA}, Article \bibinfo{articleno}{32}, \bibinfo{numpages}{17}~pages.
\newblock
\showISBNx{9798400703300}
\href{https://doi.org/10.1145/3613904.3642461}{doi:\nolinkurl{10.1145/3613904.3642461}}


\bibitem[Skjuve et~al\mbox{.}(2024)]%
        {skjuve2024people}
\bibfield{author}{\bibinfo{person}{Marita Skjuve}, \bibinfo{person}{Petter~Bae Brandtz{\ae}g}, {and} \bibinfo{person}{Asbj{\o}rn F{\o}lstad}.} \bibinfo{year}{2024}\natexlab{}.
\newblock \showarticletitle{Why do people use ChatGPT? Exploring user motivations for generative conversational AI}.
\newblock \bibinfo{journal}{\emph{First Monday}} \bibinfo{volume}{29}, \bibinfo{number}{1} (\bibinfo{year}{2024}), \bibinfo{pages}{1--23}.
\newblock


\bibitem[Skjuve et~al\mbox{.}(2021)]%
        {skjuve2021my}
\bibfield{author}{\bibinfo{person}{Marita Skjuve}, \bibinfo{person}{Asbj{\o}rn F{\o}lstad}, \bibinfo{person}{Knut~Inge Fostervold}, {and} \bibinfo{person}{Petter~Bae Brandtzaeg}.} \bibinfo{year}{2021}\natexlab{}.
\newblock \showarticletitle{My chatbot companion-a study of human-chatbot relationships}.
\newblock \bibinfo{journal}{\emph{International Journal of Human-Computer Studies}}  \bibinfo{volume}{149} (\bibinfo{year}{2021}), \bibinfo{pages}{102601}.
\newblock


\bibitem[Slota et~al\mbox{.}(2023)]%
        {slota2023many}
\bibfield{author}{\bibinfo{person}{Stephen~C. Slota}, \bibinfo{person}{Kenneth~R. Fleischmann}, \bibinfo{person}{Sherri Greenberg}, \bibinfo{person}{Nitin Verma}, \bibinfo{person}{Brenna Cummings}, \bibinfo{person}{Lan Li}, {and} \bibinfo{person}{Chris Shenefiel}.} \bibinfo{year}{2023}\natexlab{}.
\newblock \showarticletitle{Many hands make many fingers to point: challenges in creating accountable AI}.
\newblock \bibinfo{journal}{\emph{AI \& Society}}  \bibinfo{volume}{38} (\bibinfo{year}{2023}), \bibinfo{pages}{1--13}.
\newblock


\bibitem[Smithson(2000)]%
        {smithson2000using}
\bibfield{author}{\bibinfo{person}{Janet Smithson}.} \bibinfo{year}{2000}\natexlab{}.
\newblock \showarticletitle{Using and analysing focus groups: limitations and possibilities}.
\newblock \bibinfo{journal}{\emph{International journal of social research methodology}} \bibinfo{volume}{3}, \bibinfo{number}{2} (\bibinfo{year}{2000}), \bibinfo{pages}{103--119}.
\newblock


\bibitem[Star and Strauss(1999)]%
        {star1999layers}
\bibfield{author}{\bibinfo{person}{Susan~Leigh Star} {and} \bibinfo{person}{Anselm Strauss}.} \bibinfo{year}{1999}\natexlab{}.
\newblock \showarticletitle{Layers of silence, arenas of voice: The ecology of visible and invisible work}.
\newblock \bibinfo{journal}{\emph{Computer supported cooperative work (CSCW)}}  \bibinfo{volume}{8} (\bibinfo{year}{1999}), \bibinfo{pages}{9--30}.
\newblock


\bibitem[Suchman(1993)]%
        {suchman1993categories}
\bibfield{author}{\bibinfo{person}{Lucy Suchman}.} \bibinfo{year}{1993}\natexlab{}.
\newblock \showarticletitle{Do categories have politics? The language/action perspective reconsidered}.
\newblock \bibinfo{journal}{\emph{Computer supported cooperative work (CSCW)}}  \bibinfo{volume}{2} (\bibinfo{year}{1993}), \bibinfo{pages}{177--190}.
\newblock


\bibitem[Suchman(1995)]%
        {suchman1995making}
\bibfield{author}{\bibinfo{person}{Lucy Suchman}.} \bibinfo{year}{1995}\natexlab{}.
\newblock \showarticletitle{Making work visible}.
\newblock \bibinfo{journal}{\emph{Commun. ACM}} \bibinfo{volume}{38}, \bibinfo{number}{9} (\bibinfo{year}{1995}), \bibinfo{pages}{56--64}.
\newblock


\bibitem[Suchman(2023)]%
        {suchman2023uncontroversial}
\bibfield{author}{\bibinfo{person}{Lucy Suchman}.} \bibinfo{year}{2023}\natexlab{}.
\newblock \showarticletitle{The uncontroversial ‘thingness’ of AI}.
\newblock \bibinfo{journal}{\emph{Big Data \& Society}} \bibinfo{volume}{10}, \bibinfo{number}{2} (\bibinfo{year}{2023}), \bibinfo{pages}{20539517231206794}.
\newblock


\bibitem[Suchman(1987)]%
        {suchman1987plans}
\bibfield{author}{\bibinfo{person}{Lucille~Alice Suchman}.} \bibinfo{year}{1987}\natexlab{}.
\newblock \bibinfo{booktitle}{\emph{Plans and situated actions: The problem of human-machine communication}}.
\newblock \bibinfo{publisher}{Cambridge University Press}, \bibinfo{address}{Boston, MA, USA}.
\newblock


\bibitem[Toader et~al\mbox{.}(2019)]%
        {toader2019effect}
\bibfield{author}{\bibinfo{person}{Diana-Cezara Toader}, \bibinfo{person}{Grațiela Boca}, \bibinfo{person}{Rita Toader}, \bibinfo{person}{Mara M{\u{a}}celaru}, \bibinfo{person}{Cezar Toader}, \bibinfo{person}{Diana Ighian}, {and} \bibinfo{person}{Adrian~T R{\u{a}}dulescu}.} \bibinfo{year}{2019}\natexlab{}.
\newblock \showarticletitle{The effect of social presence and chatbot errors on trust}.
\newblock \bibinfo{journal}{\emph{Sustainability}} \bibinfo{volume}{12}, \bibinfo{number}{1} (\bibinfo{year}{2019}), \bibinfo{pages}{256}.
\newblock


\bibitem[Troshani et~al\mbox{.}(2021)]%
        {troshani2021we}
\bibfield{author}{\bibinfo{person}{Indrit Troshani}, \bibinfo{person}{Sally Rao~Hill}, \bibinfo{person}{Claire Sherman}, {and} \bibinfo{person}{Damien Arthur}.} \bibinfo{year}{2021}\natexlab{}.
\newblock \showarticletitle{Do we trust in AI? Role of anthropomorphism and intelligence}.
\newblock \bibinfo{journal}{\emph{Journal of Computer Information Systems}} \bibinfo{volume}{61}, \bibinfo{number}{5} (\bibinfo{year}{2021}), \bibinfo{pages}{481--491}.
\newblock


\bibitem[Tsai et~al\mbox{.}(2021)]%
        {tsai2021chatbots}
\bibfield{author}{\bibinfo{person}{Wan-Hsiu~Sunny Tsai}, \bibinfo{person}{Yu Liu}, {and} \bibinfo{person}{Ching-Hua Chuan}.} \bibinfo{year}{2021}\natexlab{}.
\newblock \showarticletitle{How chatbots' social presence communication enhances consumer engagement: the mediating role of parasocial interaction and dialogue}.
\newblock \bibinfo{journal}{\emph{Journal of Research in Interactive Marketing}} \bibinfo{volume}{15}, \bibinfo{number}{3} (\bibinfo{year}{2021}), \bibinfo{pages}{460--482}.
\newblock


\bibitem[Vaidyam et~al\mbox{.}(2019)]%
        {vaidyam2019chatbots}
\bibfield{author}{\bibinfo{person}{Aditya~Nrusimha Vaidyam}, \bibinfo{person}{Hannah Wisniewski}, \bibinfo{person}{John~David Halamka}, \bibinfo{person}{Matcheri~S Kashavan}, {and} \bibinfo{person}{John~Blake Torous}.} \bibinfo{year}{2019}\natexlab{}.
\newblock \showarticletitle{Chatbots and conversational agents in mental health: a review of the psychiatric landscape}.
\newblock \bibinfo{journal}{\emph{The Canadian Journal of Psychiatry}} \bibinfo{volume}{64}, \bibinfo{number}{7} (\bibinfo{year}{2019}), \bibinfo{pages}{456--464}.
\newblock


\bibitem[Valenzuela and Hadi(2017)]%
        {valenzuela2017implications}
\bibfield{author}{\bibinfo{person}{Ana Valenzuela} {and} \bibinfo{person}{Rhonda Hadi}.} \bibinfo{year}{2017}\natexlab{}.
\newblock \showarticletitle{Implications of product anthropomorphism through design}.
\newblock In \bibinfo{booktitle}{\emph{The Routledge Companion to Consumer Behavior}}. \bibinfo{publisher}{Routledge}, \bibinfo{address}{New York, New York, USA}, \bibinfo{pages}{82--96}.
\newblock


\bibitem[Vinsel(2023)]%
        {vinsel2023}
\bibfield{author}{\bibinfo{person}{Lee Vinsel}.} \bibinfo{year}{2023}\natexlab{}.
\newblock \bibinfo{title}{Don’t Get Distracted by the Hype Around Generative AI}.
\newblock \bibinfo{howpublished}{\url{https://sloanreview.mit.edu/article/dont-get-distracted-by-the-hype-around-generative-ai/}}.
\newblock


\bibitem[Wang(2017)]%
        {WANG2017334}
\bibfield{author}{\bibinfo{person}{Wenhuan Wang}.} \bibinfo{year}{2017}\natexlab{}.
\newblock \showarticletitle{Smartphones as Social Actors? Social dispositional factors in assessing anthropomorphism}.
\newblock \bibinfo{journal}{\emph{Computers in Human Behavior}}  \bibinfo{volume}{68} (\bibinfo{year}{2017}), \bibinfo{pages}{334--344}.
\newblock
\showISSN{0747-5632}
\href{https://doi.org/10.1016/j.chb.2016.11.022}{doi:\nolinkurl{10.1016/j.chb.2016.11.022}}


\bibitem[Waytz et~al\mbox{.}(2010a)]%
        {waytz2010sees}
\bibfield{author}{\bibinfo{person}{Adam Waytz}, \bibinfo{person}{John Cacioppo}, {and} \bibinfo{person}{Nicholas Epley}.} \bibinfo{year}{2010}\natexlab{a}.
\newblock \showarticletitle{Who sees human? The stability and importance of individual differences in anthropomorphism}.
\newblock \bibinfo{journal}{\emph{Perspectives on Psychological Science}} \bibinfo{volume}{5}, \bibinfo{number}{3} (\bibinfo{year}{2010}), \bibinfo{pages}{219--232}.
\newblock


\bibitem[Waytz et~al\mbox{.}(2010b)]%
        {waytz2010social}
\bibfield{author}{\bibinfo{person}{Adam Waytz}, \bibinfo{person}{Nicholas Epley}, {and} \bibinfo{person}{John~T. Cacioppo}.} \bibinfo{year}{2010}\natexlab{b}.
\newblock \showarticletitle{Social cognition unbound: Insights into anthropomorphism and dehumanization}.
\newblock \bibinfo{journal}{\emph{Current Directions in Psychological Science}} \bibinfo{volume}{19}, \bibinfo{number}{1} (\bibinfo{year}{2010}), \bibinfo{pages}{58--62}.
\newblock


\bibitem[Waytz et~al\mbox{.}(2014)]%
        {waytz2014mind}
\bibfield{author}{\bibinfo{person}{Adam Waytz}, \bibinfo{person}{Joy Heafner}, {and} \bibinfo{person}{Nicholas Epley}.} \bibinfo{year}{2014}\natexlab{}.
\newblock \showarticletitle{The mind in the machine: Anthropomorphism increases trust in an autonomous vehicle}.
\newblock \bibinfo{journal}{\emph{Journal of experimental social psychology}}  \bibinfo{volume}{52} (\bibinfo{year}{2014}), \bibinfo{pages}{113--117}.
\newblock


\bibitem[Weick(1995)]%
        {weick1995sensemaking}
\bibfield{author}{\bibinfo{person}{Karl~E. Weick}.} \bibinfo{year}{1995}\natexlab{}.
\newblock \bibinfo{booktitle}{\emph{Sensemaking in Organizations}}. Vol.~\bibinfo{volume}{3}.
\newblock \bibinfo{publisher}{Sage}, \bibinfo{address}{Thousand Oaks, CA, USA}.
\newblock


\bibitem[Weidinger et~al\mbox{.}(2021)]%
        {weidinger2021ethical}
\bibfield{author}{\bibinfo{person}{Laura Weidinger}, \bibinfo{person}{John Mellor}, \bibinfo{person}{Maribeth Rauh}, \bibinfo{person}{Conor Griffin}, \bibinfo{person}{Jonathan Uesato}, \bibinfo{person}{Po-Sen Huang}, \bibinfo{person}{Myra Cheng}, \bibinfo{person}{Mia Glaese}, \bibinfo{person}{Borja Balle}, \bibinfo{person}{Atoosa Kasirzadeh}, \bibinfo{person}{Zac Kenton}, \bibinfo{person}{Sasha Brown}, \bibinfo{person}{Will Hawkins}, \bibinfo{person}{Tom Stepleton}, \bibinfo{person}{Courtney Biles}, \bibinfo{person}{Abeba Birhane}, \bibinfo{person}{Julia Haas}, \bibinfo{person}{Laura Rimell}, \bibinfo{person}{Lisa~Anne Hendricks}, \bibinfo{person}{William Isaac}, \bibinfo{person}{Sean Legassick}, \bibinfo{person}{Geoffrey Irving}, {and} \bibinfo{person}{Iason Gabriel}.} \bibinfo{year}{2021}\natexlab{}.
\newblock \bibinfo{title}{Ethical and social risks of harm from Language Models}.
\newblock
\showeprint[arxiv]{2112.04359}~[cs.CL]
\urldef\tempurl%
\url{https://arxiv.org/abs/2112.04359}
\showURL{%
\tempurl}


\bibitem[Weidinger et~al\mbox{.}(2022)]%
        {weidinger2022}
\bibfield{author}{\bibinfo{person}{Laura Weidinger}, \bibinfo{person}{Jonathan Uesato}, \bibinfo{person}{Maribeth Rauh}, \bibinfo{person}{Conor Griffin}, \bibinfo{person}{Po-Sen Huang}, \bibinfo{person}{John Mellor}, \bibinfo{person}{Amelia Glaese}, \bibinfo{person}{Myra Cheng}, \bibinfo{person}{Borja Balle}, \bibinfo{person}{Atoosa Kasirzadeh}, \bibinfo{person}{Courtney Biles}, \bibinfo{person}{Sasha Brown}, \bibinfo{person}{Zac Kenton}, \bibinfo{person}{Will Hawkins}, \bibinfo{person}{Tom Stepleton}, \bibinfo{person}{Abeba Birhane}, \bibinfo{person}{Lisa~Anne Hendricks}, \bibinfo{person}{Laura Rimell}, \bibinfo{person}{William Isaac}, \bibinfo{person}{Julia Haas}, \bibinfo{person}{Sean Legassick}, \bibinfo{person}{Geoffrey Irving}, {and} \bibinfo{person}{Iason Gabriel}.} \bibinfo{year}{2022}\natexlab{}.
\newblock \showarticletitle{Taxonomy of Risks posed by Language Models}. In \bibinfo{booktitle}{\emph{Proceedings of the 2022 ACM Conference on Fairness, Accountability, and Transparency}} (Seoul, Republic of Korea) \emph{(\bibinfo{series}{FAccT '22})}. \bibinfo{publisher}{Association for Computing Machinery}, \bibinfo{address}{New York, NY, USA}, \bibinfo{pages}{214–229}.
\newblock
\showISBNx{9781450393522}
\href{https://doi.org/10.1145/3531146.3533088}{doi:\nolinkurl{10.1145/3531146.3533088}}


\bibitem[Wilkinson et~al\mbox{.}(2023)]%
        {wilkinson2023theories}
\bibfield{author}{\bibinfo{person}{Daricia Wilkinson}, \bibinfo{person}{Michael Ekstrand}, \bibinfo{person}{Janet~A. Vertesi}, {and} \bibinfo{person}{Alexandra Olteanu}.} \bibinfo{year}{2023}\natexlab{}.
\newblock \showarticletitle{Theories of Change in Responsible AI}. In \bibinfo{booktitle}{\emph{CRAFT Sessions}}.
\newblock


\bibitem[Woodruff et~al\mbox{.}(2024)]%
        {woodruff2024}
\bibfield{author}{\bibinfo{person}{Allison Woodruff}, \bibinfo{person}{Renee Shelby}, \bibinfo{person}{Patrick~Gage Kelley}, \bibinfo{person}{Steven Rousso-Schindler}, \bibinfo{person}{Jamila Smith-Loud}, {and} \bibinfo{person}{Lauren Wilcox}.} \bibinfo{year}{2024}\natexlab{}.
\newblock \showarticletitle{How Knowledge Workers Think Generative AI Will (Not) Transform Their Industries}. In \bibinfo{booktitle}{\emph{Proceedings of the 2024 CHI Conference on Human Factors in Computing Systems}} (Honolulu, HI, USA) \emph{(\bibinfo{series}{CHI '24})}. \bibinfo{publisher}{Association for Computing Machinery}, \bibinfo{address}{New York, NY, USA}, Article \bibinfo{articleno}{641}, \bibinfo{numpages}{26}~pages.
\newblock
\showISBNx{9798400703300}
\href{https://doi.org/10.1145/3613904.3642700}{doi:\nolinkurl{10.1145/3613904.3642700}}


\bibitem[Z{\l}otowski et~al\mbox{.}(2015)]%
        {zlotowski2015anthropomorphism}
\bibfield{author}{\bibinfo{person}{Jakub Z{\l}otowski}, \bibinfo{person}{Diane Proudfoot}, \bibinfo{person}{Kumar Yogeeswaran}, {and} \bibinfo{person}{Christoph Bartneck}.} \bibinfo{year}{2015}\natexlab{}.
\newblock \showarticletitle{Anthropomorphism: opportunities and challenges in human--robot interaction}.
\newblock \bibinfo{journal}{\emph{International journal of social robotics}}  \bibinfo{volume}{7} (\bibinfo{year}{2015}), \bibinfo{pages}{347--360}.
\newblock


\end{thebibliography}

\appendix
\pagebreak

\FloatBarrier
\section{Appendix}
\label{appendix}

\subsection{Focus Group Protocol}
After brief introductions, we informed workers that we aimed to understand how they navigate opportunities and challenges posed by genAI. After emphasizing that humanlike characteristics have played a role in both excitement and concern around new technologies, we provided examples of varied technologies that might be described as “humanlike” or “anthropomorphic” (e.g., car headlights that look like a sad face; a pet rock that a child might “care for”; Siri, which is voiced and interactive). In sharing examples, we highlighted different bases for why one might describe them as humanlike (i.e., visual appearance; implied interaction; social characteristics), as well as risks commonly discussed in public media (e.g., overtrust; misinformation). Participants were then asked to provide other examples they could think of that might be similar to or completely different from the ones provided.

\begin{figure}
\begin{subfigure}[b]{0.45\textwidth}   
        \centering
        \caption{}
        \includegraphics[width=\textwidth]{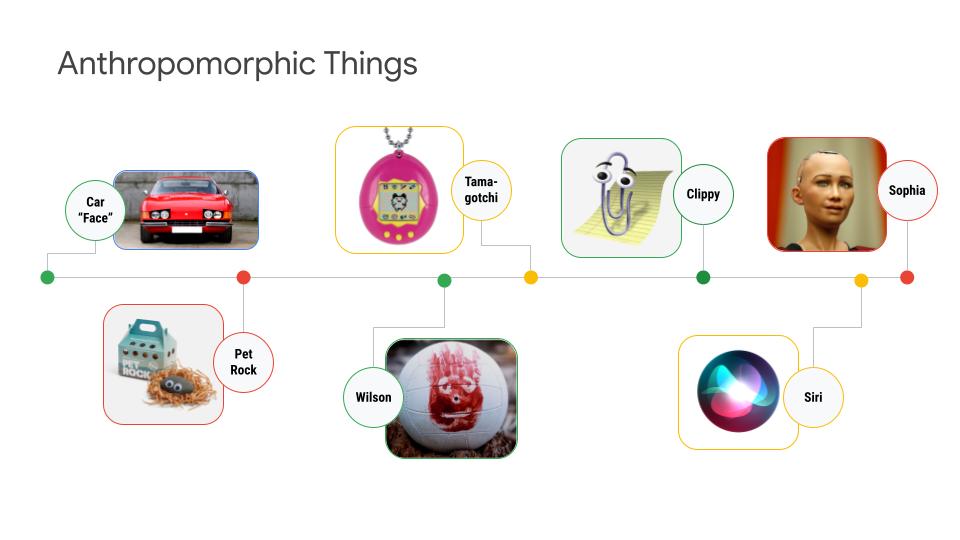}
        \Description[Presentation slide titled ``Anthropomorphic Things’’]{The slide depicts seven photos of anthropomorphic technologies including, car headlights that resemble a sad face, a `pet rock’ with googly eyes, the Wilson volleyball from the film Castaway, a Tamagotchi toy, Microsoft’s Clippy, the app icon for Siri, and a humanoid robot named Sophia.}
    \end{subfigure}
    \begin{subfigure}[b]{0.45\textwidth}
        \centering
        \caption{}
        \label{interface}
        \includegraphics[width=\textwidth]{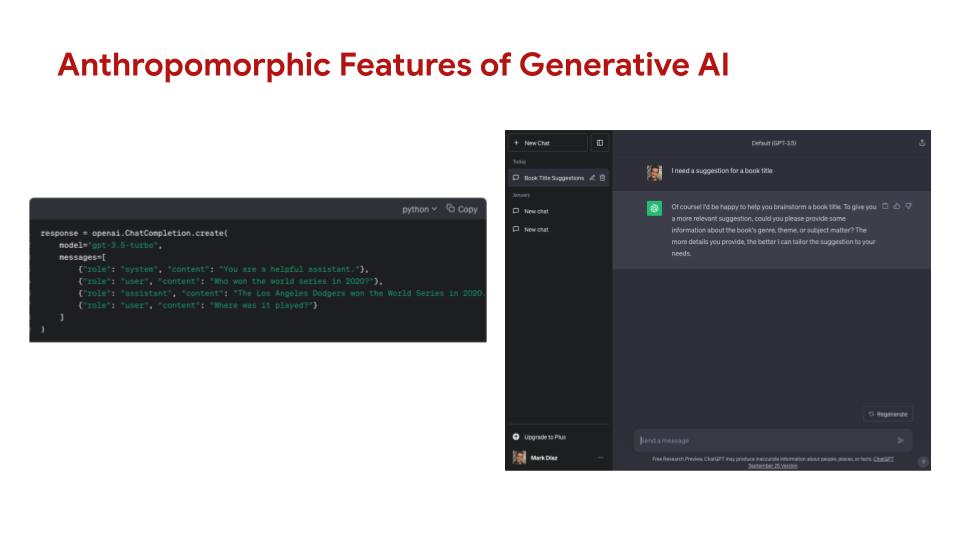} 
        \Description[Presentation slide titled ``Anthropomorphic Features of Generative AI’’]{he slide depicts a screenshot of a programming interface with written code to query one of OpenAI’s GPT models next to a second screen shot of the ChatGPT interface}
    \end{subfigure}
    \caption{The slides presented to tech workers introducing humanlikeness through examples. (a) shows examples ranging in sophistication from a ``pet rock'' to advanced AI robots. (b) shows a programming interface for querying GPT models programmatically next to the ChatGPT interface in order to highlight how a system interface can influence humanlikeness of the same underlying AI model.}
\end{figure}

Before beginning the main focus group activity, we emphasized that genAI systems are not inherently humanlike or humanlike to the same degree. To illustrate this, we offered an example of the ability to interact with OpenAI's GPT models via code compared with the ability to interact with GPT models via the ChatGPT interface (Figure \ref{interface}). While we encouraged participants to think beyond chatbot-style genAI, these technologies were a primary focus of discussion, along with image generation and non-chat based text-generation. Participants most frequently discussed these general categories of technology but also made frequent reference to ChatGPT, Bard (now named Gemini), and their associated APIs.

\begin{figure}
\begin{subfigure}[b]{0.45\textwidth}   
        \centering
        \caption{}
        \includegraphics[width=\textwidth]{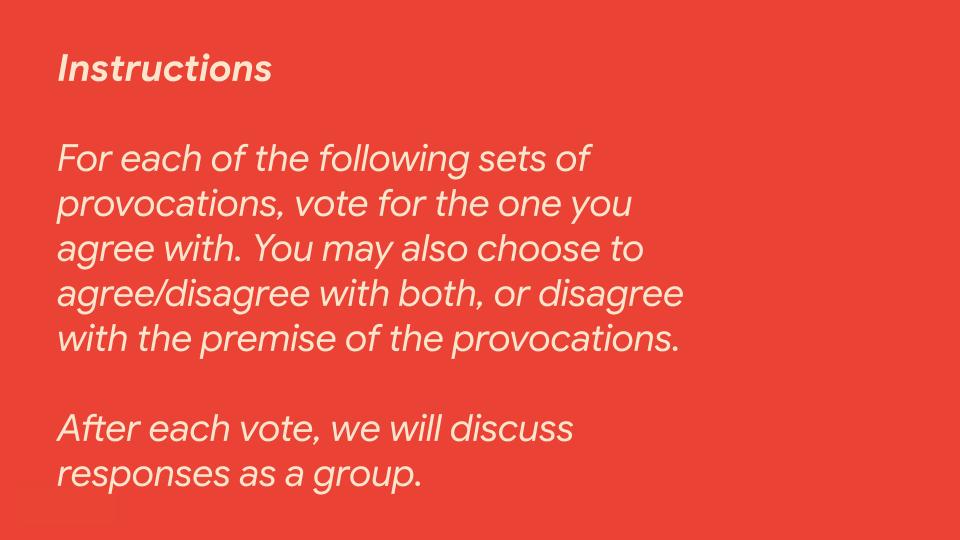}
        \Description[Presentation slide titled ``Instructions’’]{The slide text states, ``For each of the following sets of provocations, vote for the one you agree with. You may also choose to agree/disagree with both, or disagree with the premise of the provocations. After each vote, we will discuss responses as a group.’’}
    \end{subfigure}
    \begin{subfigure}[b]{0.45\textwidth}
        \centering
        \caption{}
        \includegraphics[width=\textwidth]{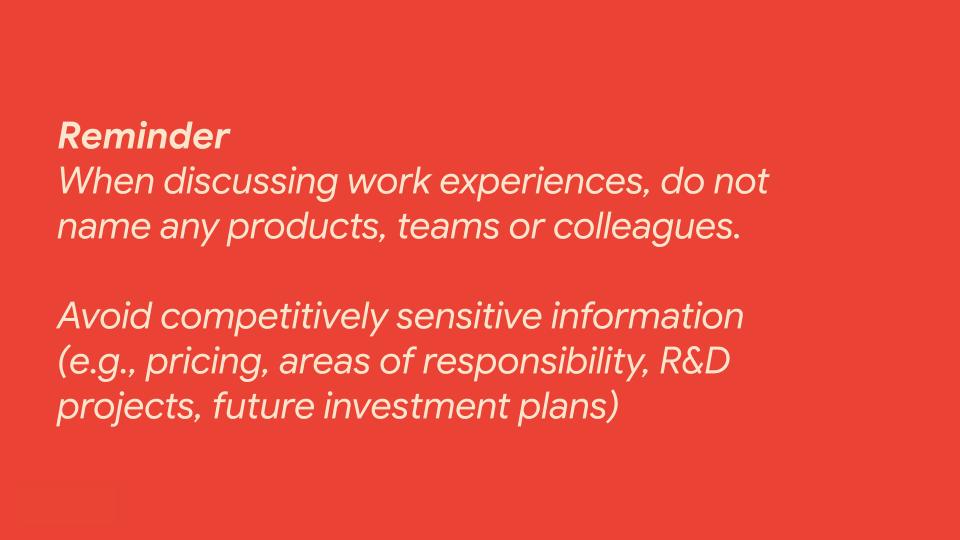} 
        \Description[Presentation slide titled ``Reminder’’]{The slide text states, ``When discussing work experiences, do not name any products, teams or colleagues. Avoid competitively sensitive information (e.g., pricing, areas of responsibility, R&D projects, future investment plans)’’}
    \end{subfigure}
    \caption{The instructions provided to tech workers before beginning the main focus group activity.}
    \label{instructions}
\end{figure}

\begin{figure}[b]
    \centering
    \includegraphics[width=0.55\textwidth]{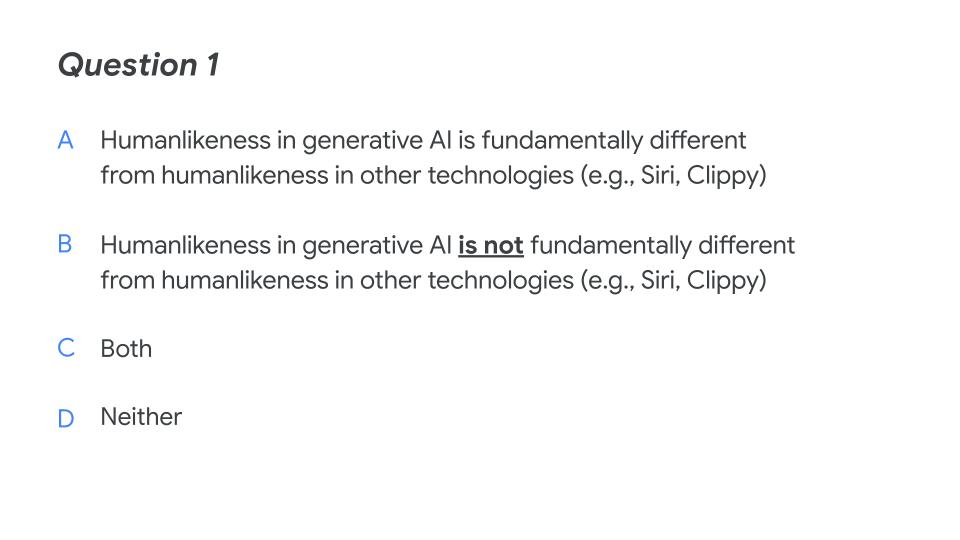}
    \Description[Presentation slide titled ``Question 1’’]{The slide depicts four options, `A. Humanlikeness in generative AI is fundamentally different from humanlikeness in other technologies (e.g., Siri, Clippy)’, `B. Humanlikeness in generative AI *is not* fundamentally different from humanlikeness in other technologies (e.g., Siri, Clippy)’, `C. Both’, `D. Neither’.}
    \caption{An example of the provocations shown. Tech workers were asked their agreement on a reductive statement related to humanlike genAI.}
    \label{example}
\end{figure}

The main focus group activity (shown in Figures \ref{instructions} and \ref{example}) consisted in structured discussion prompted by simple, reductive provocations about humanlike genAI, which workers were asked to agree or disagree with. As described in \ref{method}, workers were encouraged to add nuance to their choices as well as challenge the premises of the questions if they desired. After the six provocations were discussed, participants were offered an opportunity to share any final thoughts before the research team concluded the session.

\subsection{Focus Group Material Examples}

\begin{figure}
\begin{subfigure}[b]{0.45\textwidth}   
        \centering
        \caption{}
        \includegraphics[width=\textwidth]{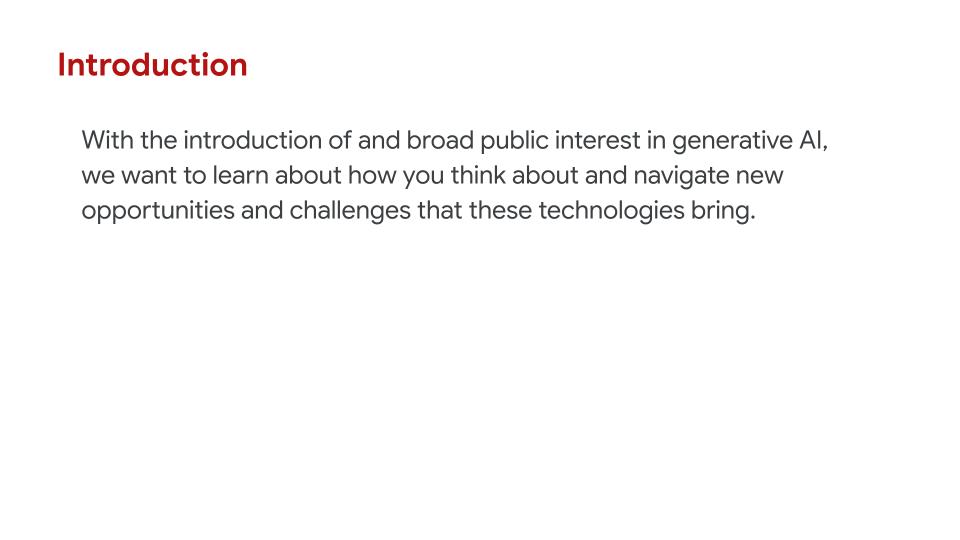}
        \Description[First introductory presentation slide titled ``Introduction’’]{The slide text states, ``With the introduction of and broad public interest in generative AI, we want to learn about how you think about and navigate new opportunities and challenges that these technologies bring.’’}
    \end{subfigure}
    \begin{subfigure}[b]{0.45\textwidth}
        \centering
        \caption{}
        \includegraphics[width=\textwidth]{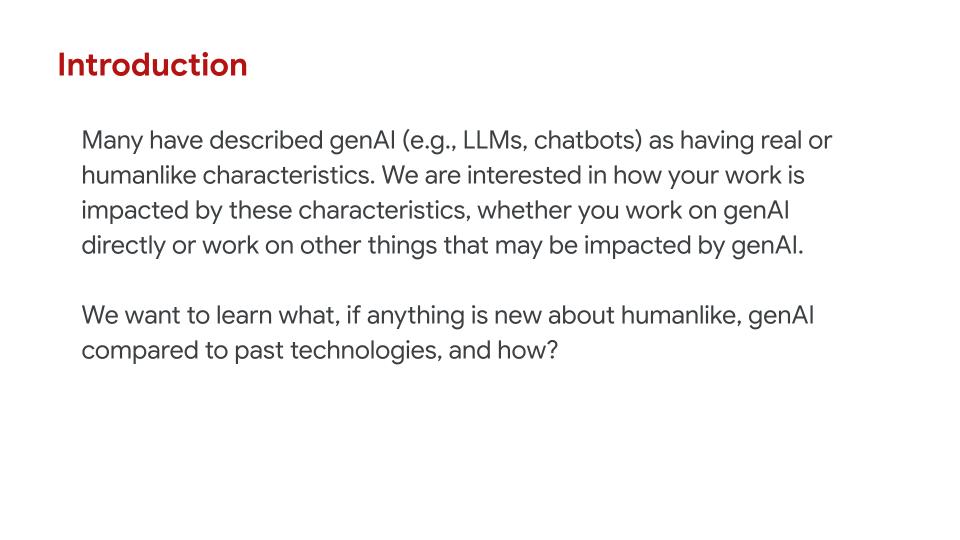} 
        \Description[Second introductory presentation slide titled ``Introduction’’]{The slide text states, ``Many have described genAI (e.g., LLMs, chatbots) as having real or humanlike characteristics. We are interested in how your work is impacted by these characteristics, whether you work on genAI directly or work on other things that may be impacted by genAI. We want to learn what, if anything is new about humanlike, genAI compared to past technologies, and how?’’}
    \end{subfigure}
    \caption{Introductory slides shown to tech workers describing the focus group themes. Humanlike genAI was introduced without specific definition.}
\end{figure}

\end{document}